\documentclass[journal,final]{IEEEtran}
\usepackage{amsmath,amsfonts}
\usepackage{amssymb}
\usepackage{amsthm}
\usepackage{algorithm}
\usepackage{array}
\usepackage[caption=false, font=footnotesize, labelfont=footnotesize, textfont=footnotesize]{subfig}
\usepackage{textcomp}
\usepackage{stfloats}
\usepackage[caption=false]{subfig}
\usepackage{url}
\usepackage{verbatim}
\usepackage{graphicx}
\usepackage{cite}
\usepackage{multirow}
\usepackage{algorithmicx}
\usepackage{algpseudocode}
\usepackage{bm} 
\usepackage{booktabs}
\usepackage{xcolor}

\usepackage{graphicx}    

\begin{document}
\bstctlcite{IEEEexample:BSTcontrol}

\title{A Gram-Attention Learning Framework for Spatially Non-Stationary Channel Estimation}

\author{Jiannan~Wang, Xianghao~Yu,~\IEEEmembership{Senior Member,~IEEE}, Chenshu~Wu,~\IEEEmembership{Senior Member,~IEEE}
    \thanks{
    Jiannan Wang and Xianghao Yu are with the Department of Electrical Engineering, City University of Hong Kong, Hong Kong (email: jiannan.wang@my.cityu.edu.hk, alex.yu@cityu.edu.hk). \textit{(Corresponding Author: Xianghao Yu)}
    
    Chenshu Wu is with the Department of Computer Science, The University of Hong Kong, Hong Kong (e-mail: chenshu@cs.hku.hk).
    }
}

\maketitle

\begin{abstract}
As next-generation wireless systems migrate towards the high-frequency bands, extremely large-scale multiple-input multiple-output (XL-MIMO) is indispensable for combating severe path loss. However, the received energy of a propagation path along the extremely large array aperture (ELAA) may exhibit spatial non-stationarity, rendering the conventional discrete Fourier transform (DFT) codebook inadequate for channel estimation (CE) due to its full-array angular representation. To overcome this model mismatch, we propose a novel joint angle-subarray (JAS) codebook, where each codeword is associated with an angular direction and a specific subarray. In this context, spatially non-stationary CE is transformed into the structured sparse support detection in the JAS domain. Nevertheless, achieving reliable support recovery remains challenging for conventional model-based methods. To address this issue, we further develop a signal processing and deep learning co-design framework, where data-driven JAS parallel support detection is followed by least-squares (LS)-based channel reconstruction. Specifically, the matched filter response of each JAS grid is first embedded into a high-dimensional feature vector. Then, we propose a Gram-attention module that incorporates the Gram matrix as a physical prior to guide feature purification, enabling the network to suppress leakage-related components while preserving support-relevant information of each grid. The purified features are further passed to decoupled heads to infer the hierarchical sparse supports in the JAS domain. Finally, LS is performed over the low-dimensional dictionary constructed from the detected supports to retrieve the path gains. Simulation results demonstrate that the proposed framework achieves the highest support detection F1-score and the lowest CE normalized mean square error (NMSE) among all baselines, while ablation studies verify the benefits of integrating the Gram-domain physical prior with the data-driven representation learning.
\end{abstract}

\begin{IEEEkeywords}
Channel estimation, deep learning, joint angle-subarray codebook, spatially non-stationary, support detection.
\end{IEEEkeywords}

\section{Introduction}
The evolution towards the sixth-generation (6G) wireless communication systems is driven by the growing demand for ultra-high data rates, ultra-low latency, and massive connectivity~\cite{9144301,10054381,9349624}. To meet these requirements, a promising solution is to exploit the high-frequency bands, such as millimeter-wave (mmWave) and terahertz (THz) bands, where abundant spectrum resources are available~\cite{10494372,9887921,wang2026throughput}. However, high-frequency propagation suffers from severe path loss, resulting in a significant degradation of the received signal-to-noise ratio (SNR). To compensate for this loss, extremely large-scale multiple-input multiple-output (XL-MIMO) systems have been widely recognized as a key enabling technology, providing high directional beamforming gains through a substantially enlarged array aperture~\cite{7400949,10379539}. In this context, acquiring accurate channel state information (CSI) is essential for fully exploiting the potential of XL-MIMO, positioning robust and reliable channel estimation (CE) as a fundamental prerequisite.

Nevertheless, due to the massive number of antennas deployed at the base station (BS), assigning a dedicated and power-hungry radio-frequency (RF) chain to each antenna is prohibitively expensive. In this regard, the hybrid architecture becomes a practical alternative, where a limited number of RF chains are connected to a large antenna array through hybrid combining networks~\cite{7397861,7445130,6847111}. However, since the number of RF chains is typically much smaller than the number of antennas, conventional least-squares (LS)-based CE incurs heavy pilot overhead to recover the high-dimensional channel, which reduces the effective throughput during the subsequent data transmission stage~\cite{10845870}. To alleviate this overhead, compressed sensing (CS)-based methods have been widely adopted for CE by exploiting the angular-domain sparsity of the high-frequency channel, typically based on the classical discrete Fourier transform (DFT) codebook~\cite{7458188}. However, the DFT codebook implicitly assumes full-array visibility for each propagation path. For an ELAA spanning tens of wavelengths or more, a path may illuminate only part of the array, causing DFT representation mismatch and thereby energy leakage in the angular domain~\cite{9170651}.

Recent efforts have investigated spatially non-stationary CE from different perspectives. Some works decomposed the original non-stationary channel into several more tractable local estimation problems. For instance, group time block code (GTBC)-based schemes designed time-block pilots to separate the spatially non-stationary channel into multiple stationary subchannels, over which sparse recovery was performed using polar-domain or adaptive codebooks~\cite{10373799,10631699}. Following the local stationarity principle, subarray-wise methods estimated visible scatterers from each local array, while scatterer-wise methods further identified the visibility regions associated with individual scatterers~\cite{8949454}. To better accommodate irregular visibility patterns, adaptive subarray partitioning was also studied in ~\cite{yang2025}, where the array segmentation was adjusted according to the spatial power distribution. Other works incorporated structured priors or learning-based models into the estimation procedure. For example, hidden Markov model (HMM)-based sparsity priors were used to describe the visibility evolution along the array aperture, leading to Bayesian inference algorithms such as turbo orthogonal approximate message passing (Turbo-OAMP)~\cite{9547795}. Model-driven deep learning was explored in~\cite{9110882}, where the channel was treated as an image and the object detection network was applied to estimate angles, delays, and visibility regions.

Despite these advances, sparse representation for spatially non-stationary channel remains less explored. To fill this gap, we develop a novel visibility-aware sparse representation inspired by the localization principle of the classical short-time Fourier transform (STFT) in radar signal processing~\cite{670330}. The key idea is to extend the conventional codebook from a one-dimensional (1D) angle-based representation to a two-dimensional (2D) angle-subarray representation. Specifically, we propose a joint angle-subarray (JAS) codebook, where each codeword corresponds to an angular direction and a specific subarray. By jointly encoding the path direction and the local visibility, the proposed codebook mitigates the representation mismatch caused by the conventional DFT codebook under spatial non-stationarity. As a result, spatially non-stationary CE can be reformulated as a structured sparse recovery problem in the JAS domain.

However, reliable support detection over the JAS codebook remains challenging for conventional methods. For example, greedy pursuit methods such as orthogonal matching pursuit (OMP) rely on serial correlation-based codeword selection, making them vulnerable to error propagation once a wrong support is identified~\cite{342465,4385788}. Sparse-group LASSO (SGL)-type formulations provide a parallel sparse optimization alternative by jointly estimating all JAS coefficients while imposing hierarchical sparsity~\cite{simon2013sparse}, but their fixed shrinkage rules are not sufficiently adaptive to distinguish true supports from neighboring leakage responses. Consequently, relying solely on the model-driven sparse recovery may limit the achievable CE accuracy under the JAS representation. Therefore, a more adaptive mechanism is required to better exploit the visibility-aware JAS representation, enabling more reliable support detection and improved CE accuracy.

To realize this mechanism, we develop a signal processing and deep learning co-design framework. Rather than regressing the dense channel directly, the learning module focuses on identifying the active supports over the JAS codebook. Then, given the detected supports, the original underdetermined CE problem is converted into an overdetermined gain estimation problem over a low-dimensional dictionary, which can be well solved by the LS method. This separation allows the neural network to address the most challenging part of the problem, namely parallel support detection over the JAS grids, while retaining a model-based estimator for continuous gain recovery. To further improve the reliability of support inference, the physical prior embedded in the Gram matrix is incorporated into the neural detector. The main contributions of this paper can be summarized as follows:
\begin{enumerate}
    \item \textbf{JAS Codebook:} We propose a novel JAS codebook for spatially non-stationary channels, whose codeword is constructed by windowing an angular steering vector over a specific subarray. Furthermore, we analyze the column coherence of the JAS codebook, revealing its structured angular-subarray correlation and the limitations of conventional algorithms over this codebook.

    \item \textbf{Two-Stage Framework:} We establish a two-stage CE procedure based on the proposed JAS representation, where active JAS supports are first detected and then used to construct a reduced dictionary for LS-based gain estimation. This formulation casts the key inference step as parallel multi-label support classification over the JAS grids, thereby exploiting the complementary strengths of data-driven classification and model-based estimation.

    \item \textbf{Gram-Attention Learning:} We develop a Gram-guided deep learning model to enable adaptive parallel support inference over the JAS codebook. Specifically, the matched filter response of each JAS grid is first embedded into a high-dimensional feature vector, and the Gram matrix prior is injected into the attention operation to guide successive feature refinement. In this way, the network learns to suppress leakage-related components and produces purified features for reliable support detection in the JAS domain.

    \item \textbf{Performance Verification:} We validate the proposed framework through comprehensive simulations and ablation studies. The results demonstrate that the proposed method improves both JAS support detection and final channel reconstruction accuracy over eight representative model-based and learning-based baselines. The ablation studies further verify the importance of combining the physical prior encoded in the Gram matrix with the data-driven representation learning for support detection.
\end{enumerate}

The remainder of this paper is organized as follows. Section~\ref{section: system model} presents the system model and the spatially non-stationary channel model. Section~\ref{sec: codebook} introduces the proposed JAS codebook and discusses the motivations for data-driven parallel inference. Section~\ref{sec: learning} details the proposed two-stage CE framework. Section~\ref{sec: simulations} presents the simulation results and ablation studies. Finally, Section~\ref{sec: conclusion} concludes this paper.

\textit{Notations}: Scalars, vectors, and matrices are denoted by lower-case $x$, bold lower-case $\mathbf{x}$, and bold upper-case $\mathbf{X}$, respectively. The transpose and conjugate transpose of $\mathbf{x}$ are denoted by $\mathbf{x}^\mathrm{T}$ and $\mathbf{x}^\mathrm{H}$, respectively. The notation $\mathcal{CN}(\boldsymbol{\mu}, \mathbf{\Sigma})$ represents the complex circularly symmetric Gaussian distribution with mean $\boldsymbol{\mu}$ and covariance matrix $\mathbf{\Sigma}$. Besides, $\mathbb{R}$ and $\mathbb{C}$ denote the sets of real and complex numbers, respectively, and $\jmath = \sqrt{-1}$ denotes the imaginary unit. The operators $\odot$ and $\otimes$ represent the Hadamard and Kronecker products, respectively. For a complex scalar $x$, its real part, imaginary part, magnitude, and phase are denoted by $\Re\{x\}$, $\Im\{x\}$, $|x|$, and $\angle x$, respectively. Finally, $\mathrm{diag}(\cdot)$ and $\mathrm{blkdiag}(\cdot)$ denote the operators that construct a diagonal matrix and a block diagonal matrix from their arguments, respectively, and $\mathbf{1}_N$ represents the $N$-dimensional all-ones vector.

\section{System Model}
\label{section: system model}
This section presents the signal model for CE and then introduces the spatially non-stationary channel model. Based on this model, we further discuss why the conventional DFT-based sparse representation becomes mismatched.

\subsection{Signal Model}
In this paper, we consider an XL-MIMO system operating in the time-division duplexing (TDD) mode, where a single-antenna user equipment (UE) communicates with a BS. The BS is equipped with an $N$-antenna uniform linear array (ULA) with half-wavelength spacing. Since $N$ is typically very large in XL-MIMO systems, the fully-connected hybrid combining structure is employed at the BS to reduce the hardware cost and energy consumption~\cite{7397861}. Specifically, during the uplink pilot training phase, a total of $Q$ time slots are utilized for CE, with the estimated channel subsequently used for downlink transmission by exploiting channel reciprocity. At the $q$-th time slot, the combining matrix is given by $\mathbf{A}_q=\mathbf{W}_{\mathrm{BB},q}\mathbf{W}_{\mathrm{RF},q}\in\mathbb{C}^{N_{\mathrm{RF}}\times N}$, where $\mathbf{W}_{\mathrm{RF},q}\in\mathbb{C}^{N_{\mathrm{RF}}\times N}$ and $\mathbf{W}_{\mathrm{BB},q}\in\mathbb{C}^{N_{\mathrm{RF}}\times N_{\mathrm{RF}}}$ denote the analog and digital combining matrices, respectively, and $N_{\mathrm{RF}}$ is the number of RF chains. Besides, since the analog combiner is usually implemented by a phase shifter network, its elements are inherently subject to the constant modulus constraint~\cite{7397861}.

At the $q$-th time slot, the UE transmits a pilot symbol $s_{q} \in \mathbb{C}$ to the BS. Then, the received pilots after hybrid combining, denoted as $\mathbf{y}_{q} \in \mathbb{C}^{N_{\text{RF}} \times 1}$, are given by
\begin{equation}
    \mathbf{y}_{q} = \mathbf{A}_q \mathbf{h} s_{q} + \mathbf{A}_q \mathbf{n}_{q},
\label{eq:linear-equation}
\end{equation}
where $\mathbf{h} \in \mathbb{C}^{N \times 1}$ denotes the channel vector between the UE and the BS, and $\mathbf{n}_{q} \sim \mathcal{CN}(\mathbf{0}, \sigma_{\mathrm{n}}^2 \mathbf{I}_N)$ is the complex additive white Gaussian noise (AWGN) of the $q$-th time slot. Without loss of generality, the pilot symbols are normalized as $s_q=1$ for $q=1,2,\cdots, Q$. Then, by stacking the pilots of $Q$ time slots together, a compact form for the received pilots is obtained as
\begin{equation}
    \mathbf{y} = \mathbf{A} \mathbf{h} +  \mathbf{n},
\label{eq:stack-linear-equation}
\end{equation}
where $\mathbf{y} =[\mathbf{y}_{1}^{\mathrm{T}},\mathbf{y}_{2}^{\mathrm{T}},\cdots,\mathbf{y}_{Q}^{\mathrm{T}}]^{\mathrm{T}}$, $\mathbf{A}=[\mathbf{A}_1^{\mathrm{T}},\mathbf{A}_2^{\mathrm{T}},\cdots,\mathbf{A}_Q^{\mathrm{T}}]^{\mathrm{T}}$, and $\mathbf{n}=[\mathbf{n}_{1}^{\mathrm{T}} \mathbf{A}_1^{\mathrm{T}}, \mathbf{n}_{2}^{\mathrm{T}} \mathbf{A}_2^{\mathrm{T}},\cdots, \mathbf{n}_{Q}^{\mathrm{T}} \mathbf{A}_Q^{\mathrm{T}}]^{\mathrm{T}}$. Crucially, due to the combining matrix $\mathbf{A}$, the stacked noise vector $\mathbf{n}$ becomes colored with distribution $\mathcal{CN}(\mathbf{0}, \mathbf{C})$, where the covariance matrix $\mathbf{C}$ is given by
\begin{equation}
    \mathbf{C} = \sigma_{\mathrm{n}}^2 \mathrm{blkdiag}\left( \mathbf{A}_1\mathbf{A}_1^{\mathrm{H}}, \, \mathbf{A}_2\mathbf{A}_2^{\mathrm{H}}, \, \cdots, \, \mathbf{A}_Q\mathbf{A}_Q^{\mathrm{H}} \right)  \in \mathbb{C}^{M \times M}.
\label{eq:noise_covariance_matrix}
\end{equation}
Here, $M=QN_{\text{RF}}$ denotes the number of measurements. In conventional linear estimation, reliable recovery of $\mathbf{h}$ usually requires $M$ to be no smaller than the number of unknown channel coefficients $N$. However, this requirement can incur prohibitive pilot overhead in XL-MIMO systems. To overcome this challenge, the inherent sparsity of the wireless channel is exploited, as detailed in the following two subsections.

\subsection{Non-Stationary Channel Model}
\begin{figure}[!t]
    \centering
    \includegraphics[width=0.3\textwidth]{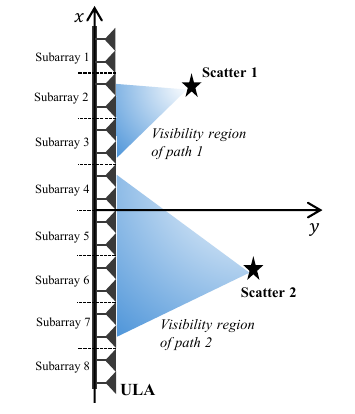} 
    \caption{Illustration of a spatially non-stationary channel with $L=2$ and $N_{\mathrm{sub}}=8$. The shaded areas denote the visibility regions of two propagation paths, corresponding to $\boldsymbol{\phi}_1=[0,1,1,0,0,0,0,0]^{\mathrm{T}}$ and $\boldsymbol{\phi}_2=[0,0,0,1,1,1,1,0]^{\mathrm{T}}$, respectively.}
    \label{fig:ns_channel}
\end{figure}

In the multi-path channel model, suppose that the spatial angle of the $l$-th path with respect to the ULA is denoted by $\theta_l \in [-1,1]$. Under the far-field assumption, the corresponding steering vector is given by
\begin{equation}
    \mathbf{b}(\theta_l) = \frac{1}{\sqrt{N}}\left[
    1,
    e^{-\jmath \frac{2\pi}{\lambda} d \theta_l},
    \cdots,
    e^{-\jmath \frac{2\pi}{\lambda} (N-1) d \theta_l}
    \right]^{\mathrm{T}},
\label{eq:far-field}
\end{equation}
where $d=\lambda/2$ is the antenna spacing and $\lambda$ denotes the wavelength. In conventional spatially stationary channel, each path is assumed to illuminate the entire array. However, different portions of an ELAA may experience different propagation environments involving distinct scattering clusters and/or obstacles. Consequently, a propagation path may be visible only over a portion of the array, as illustrated in Fig.~\ref{fig:ns_channel}.

To model this spatial non-stationarity in a tractable manner, we adopt the subarray-level visibility model in this paper~\cite{10373799,10631699,8949454}. Specifically, the BS array is divided into $N_{\mathrm{sub}}$ subarrays, each containing $N_{\mathrm b}=N/N_{\mathrm{sub}}$ antennas, where $N_{\mathrm b}$ is assumed to be an integer. Since each subarray has a much smaller aperture than the whole array, the path visibility can be approximated as locally unchanged within one subarray. Therefore, for a given path, all antennas in a specific subarray are assumed to share the same visibility state. Let $\boldsymbol{\phi}_l\in\{0,1\}^{N_{\mathrm{sub}}\times 1}$ denote the subarray-level visibility indicator of the $l$-th path, whose $i$-th element equals one if the $i$-th subarray is visible to this path and equals zero otherwise. Then, the spatially non-stationary channel is modeled as
\begin{equation}
    \mathbf{h} = \sqrt{\frac{N}{L}}
    \sum_{l=1}^L \beta_l \mathbf{b}(\theta_l) \odot
    \left(\boldsymbol{\phi}_l \otimes \mathbf{1}_{N_{\mathrm b}}\right),
\label{eq:vr-multi-path-channel}
\end{equation}
where $\beta_l$ denotes the complex gain of the $l$-th path and $L$ is the number of paths. Note that when all paths are visible to the entire array, namely $\boldsymbol{\phi}_l=\mathbf{1}_{N_{\mathrm{sub}}}$ for $l=1,2,\cdots, L$, the model in \eqref{eq:vr-multi-path-channel} reduces to the conventional spatially stationary case.

\subsection{DFT-Based Sparse Representation and Model Mismatch}
In conventional spatially stationary CE, the DFT codebook has been commonly used to exploit the sparsity of the wireless channel in the angular domain. Specifically, each codeword of the codebook matrix is a steering vector $\mathbf{b}(\bar{\theta}_g)$ sampled at the discrete angle $\bar{\theta}_g=\frac{2g-G+1}{G}$, where $g=0,1,\cdots,G-1$ and $G$ is the number of sampled angles. In this regard, the channel can be transformed from the spatial domain to the angular domain as
\begin{equation}
    \mathbf{h} = \mathbf{F}_{\mathrm{DFT}}\mathbf{h}_{\mathrm{DFT}},
\label{eq:DFT}
\end{equation}
where $\mathbf{F}_{\mathrm{DFT}}=[\mathbf{b}(\bar{\theta}_0),\mathbf{b}(\bar{\theta}_1),\cdots,\mathbf{b}(\bar{\theta}_{G-1})] \in \mathbb{C}^{N \times G}$ denotes the DFT codebook, and $\mathbf{h}_{\mathrm{DFT}}\in\mathbb{C}^{G\times 1}$ is the corresponding sparse representation in the angular domain. Substituting \eqref{eq:DFT} into \eqref{eq:stack-linear-equation}, the DFT-based measurement model is given by
\begin{equation}
    \mathbf{y} = \mathbf{A}\mathbf{F}_{\mathrm{DFT}}\mathbf{h}_{\mathrm{DFT}}+\mathbf{n}.
\label{eq:dft_measurement}
\end{equation}

Crucially, the effectiveness of this DFT-based sparse representation relies on the assumption that each path component can be well represented by the full-array steering vector $\mathbf{b}(\theta_l)$. However, according to the spatially non-stationary model in \eqref{eq:vr-multi-path-channel}, the $l$-th path is characterized by $\mathbf{b}(\theta_l)\odot(\boldsymbol{\phi}_l\otimes\mathbf{1}_{N_{\mathrm b}})$, where the angular response is further modulated by the subarray-level visibility pattern $\boldsymbol{\phi}_l$. Therefore, the conventional DFT codebook suffers from a representation mismatch, which introduces energy leakage in the angular domain and thus degrades the reliability of sparse recovery. To address this issue, we propose a novel codebook tailored to the spatially non-stationary channel in the next section.

\section{Proposed Codebook}
\label{sec: codebook}
This section introduces the principle of the proposed JAS codebook for mitigating the representation mismatch caused by spatial non-stationarity. Then, we analyze its column coherence to discuss the limitations of conventional methods.

\subsection{Proposed JAS Codebook}
\label{sec: jas}
As discussed in the last section, the spatial non-stationarity of the channel introduces a structural mismatch with the classical DFT codebook, leading to energy leakage in the angular domain. To address this issue, we construct a windowed angular representation inspired by STFT, where each full-array steering vector is localized to the individual subarray. Let $\mathbf{w}_s=\mathbf{e}_s\otimes\mathbf{1}_{N_{\mathrm b}}$ denote the binary window selecting the $s$-th subarray, where $\mathbf{e}_s\in\mathbb{R}^{N_{\mathrm{sub}}\times 1}$ is a one-hot vector whose $s$-th element is one. Then, the JAS codeword associated with the $g$-th sampled angle and the $s$-th subarray is defined as
\begin{equation}
\label{eq:jas_codeword}
    \mathbf{f}_{g,s} = \mathbf{b}(\bar{\theta}_g) \odot \mathbf{w}_s,
\end{equation}
where $g = 1,2, \cdots, G$ and $s=1, 2, \cdots, N_{\mathrm{sub}}$. In this representation, a path visible to multiple subarrays can be described by activating several JAS codewords with the same angular index but different subarray indices. Therefore, the spatial non-stationarity is well embedded into the support pattern of the JAS representation. Collecting all such angle-subarray pairs yields the proposed codebook $\mathbf{F}_{\mathrm{JAS}}\in \mathbb{C}^{N\times J}$ in \eqref{eq:jas-codebook}, which is presented on the top of the next page, where $J=GN_{\mathrm{sub}}$ denotes the size of the JAS codebook.

\begin{figure*}[t]
\begin{equation}
    \mathbf{F}_{\mathrm{JAS}} = \big[ 
    \underbrace{\mathbf{f}_{1,1}, \mathbf{f}_{1,2}, \cdots, \mathbf{f}_{1,N_{\mathrm{sub}}}}_{\text{Angle }\bar{\theta}_1}, \,
    \underbrace{\mathbf{f}_{2,1}, \mathbf{f}_{2,2}, \cdots, \mathbf{f}_{2,N_{\mathrm{sub}}}}_{\text{Angle }\bar{\theta}_2}, \,
    \cdots, \,
    \underbrace{\mathbf{f}_{G,1}, \mathbf{f}_{G,2}, \cdots, \mathbf{f}_{G,N_{\mathrm{sub}}}}_{\text{Angle }\bar{\theta}_G} 
    \big]
\label{eq:jas-codebook}
\end{equation}
\hrulefill 
\end{figure*}

In the CS-based sparse signal recovery framework, the mutual coherence of the codebook is an important metric that dictates recovery accuracy. Based on the definition in \eqref{eq:jas_codeword}, the normalized inner product between two arbitrary codewords $\mathbf{f}_{g,s}$ and $\mathbf{f}_{g',s'}$ is derived as
\begin{equation}
\begin{aligned}
    \frac{
    \left|\mathbf{f}_{g,s}^{\mathrm{H}}\mathbf{f}_{g',s'}\right|
    }{
    \|\mathbf{f}_{g,s}\|_2\|\mathbf{f}_{g',s'}\|_2
    } 
    &= N_\mathrm{sub} \left|\left(\mathbf{b}(\bar{\theta}_{g})\odot\mathbf{w}_{s}\right)^{\mathrm{H}} \left(\mathbf{b}(\bar{\theta}_{g'})\odot\mathbf{w}_{s'}\right) \right| \\
    &= \frac{1}{N_\mathrm{b}} \left| \sum_{n=0}^{N-1} [\mathbf{w}_{s}]_n [\mathbf{w}_{s'}]_n e^{\jmath \pi n (\bar{\theta}_{g} - \bar{\theta}_{g'})} \right|.
\end{aligned}
\label{eq:general_inner_product}
\end{equation}

Based on this formulation, the mutual coherence analysis is conducted from the following two distinct perspectives.

\subsubsection{Inter-Subarray Coherence ($s \neq s'$)}
When the two codewords are sampled from different subarrays, their spatial windowing vectors $\mathbf{w}_s$ and $\mathbf{w}_{s'}$ have disjoint non-zero supports, yielding $[\mathbf{w}_{s}]_n [\mathbf{w}_{s'}]_n = 0$ for all $n$. Substituting this property into \eqref{eq:general_inner_product} yields
\begin{equation}
    \frac{
    \left|\mathbf{f}_{g,s}^{\mathrm{H}}\mathbf{f}_{g',s'}\right|
    }{
    \|\mathbf{f}_{g,s}\|_2\|\mathbf{f}_{g',s'}\|_2
    }  = 0, \quad \forall s \neq s',
\end{equation}
which demonstrates that the codewords corresponding to different subarrays are strictly orthogonal to each other, regardless of their respective sampling angles. Consequently, the response with respect to a specific subarray will not introduce energy leakage to other subarrays.

\subsubsection{Intra-Subarray Coherence ($s = s'$)}
When the two codewords correspond to the same subarray but different sampled angles, i.e., $g \neq g'$, the non-zero support simplifies to the antenna index range of the $s$-th subarray. In this regard, the coherence function in \eqref{eq:general_inner_product} simplifies to
\begin{equation}
\begin{aligned}
    \frac{
    \left|\mathbf{f}_{g,s}^{\mathrm{H}}\mathbf{f}_{g',s'}\right|
    }{
    \|\mathbf{f}_{g,s}\|_2\|\mathbf{f}_{g',s'}\|_2
    }  
    &= \frac{1}{N_\mathrm{b}} \left| \sum_{n=(s-1){N_{\mathrm{b}}}}^{sN_{\mathrm{b}}-1} e^{\jmath \pi n (\bar{\theta}_{g} - \bar{\theta}_{g'}) } \right| \\
    &= \frac{1}{N_\mathrm{b}} \left| \frac{\sin\left(\frac{\pi}{2}N_{\mathrm{b}} (\bar{\theta}_{g} - \bar{\theta}_{g'}) \right)}{\sin\left(\frac{\pi}{2}(\bar{\theta}_{g} - \bar{\theta}_{g'})\right)} \right|.
\label{eq:same_subarray}
\end{aligned}
\end{equation}

As can be observed from \eqref{eq:same_subarray}, the column coherence along the angular dimension is described by a normalized Dirichlet kernel, which reaches its maximum value of one when $\bar{\theta}_{g}=\bar{\theta}_{g'}$ and then decreases as the angular separation $|\bar{\theta}_{g}-\bar{\theta}_{g'}|$ increases. Therefore, the proposed JAS dictionary exhibits a structured correlation pattern: Codewords from different subarrays are orthogonal, while codewords within the same subarray remain correlated along the angular dimension due to the finite aperture.

\subsection{From Serial Pursuit to Adaptive Parallel Inference}
\label{se: motivation}
\begin{figure}[t!]
    \centering
    \subfloat[]{
        \includegraphics[width=0.45\columnwidth]{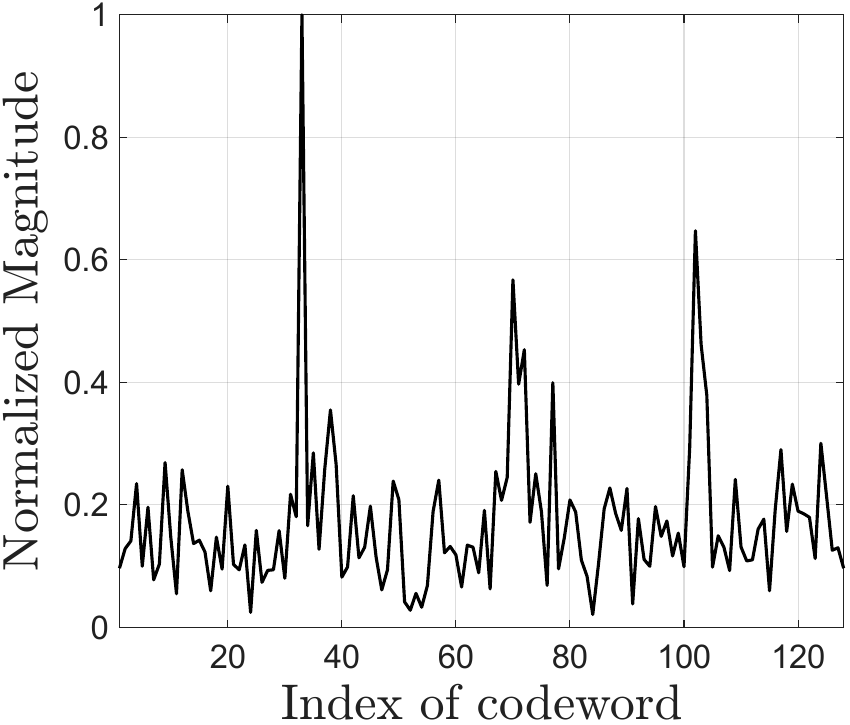}
        \label{fig:omp_dft_demo}
    }
    \hfil
    \subfloat[]{
        \includegraphics[width=0.45\columnwidth]{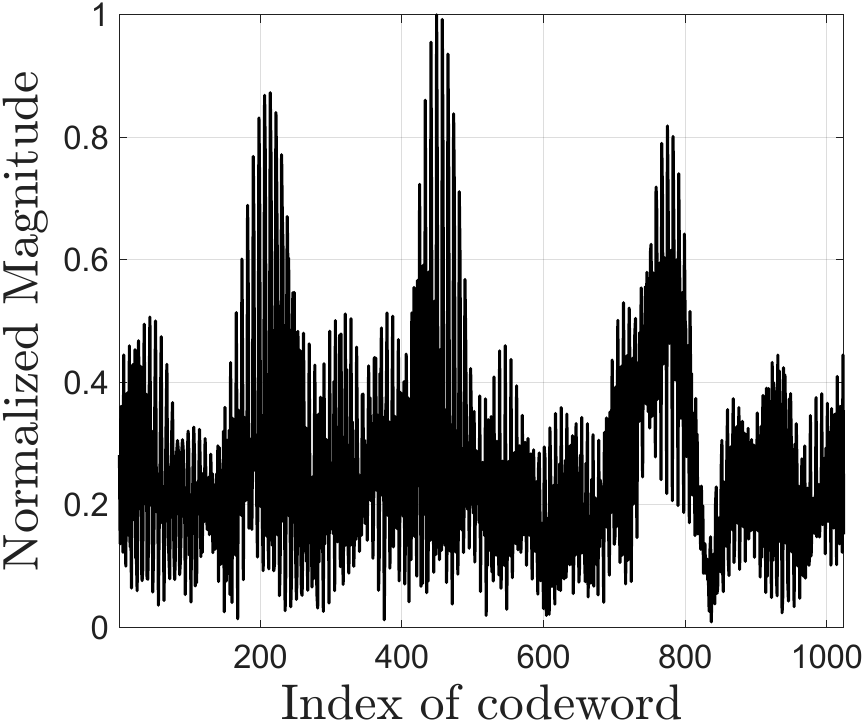}
        \label{fig:omp_jas_demo}
    }
    \\
    \caption{Magnitude distributions of correlations over the spatially non-stationary channel with three paths. The spatial angles of these three paths are configured as $-0.5$, $0.1$, and $0.6$, respectively. Besides, path 1 is visible to all subarrays, path 2 is only visible to the first $4$ subarrays while path 3 is visible to the last $4$ subarrays. (a) Normalized distribution of $|(\mathbf{A F_{\mathrm{DFT}}})^{\mathrm{H}} \mathbf{y}|$ with $G=128$. (b) Normalized distribution of $|(\mathbf{A F_{\mathrm{JAS}}})^{\mathrm{H}} \mathbf{y}|$ with $G=128$ and $N_{\mathrm{sub}}=8$.}
    \label{fig:codebook_comparison}
\end{figure}

With the proposed JAS codebook, the spatially non-stationary channel can be represented in a visibility-aware angular domain, where each active coefficient corresponds to a specific angle-subarray pair. Accordingly, the channel vector is transformed as
\begin{equation}
    \mathbf{h} = \mathbf{F}_{\mathrm{JAS}} \mathbf{h}_{\mathrm{JAS}},
\end{equation}
where $\mathbf{h}_{\mathrm{JAS}} \in \mathbb{C}^{J \times 1}$ denotes the sparse representation in the JAS domain. Based on this new representation, the measurement equation in \eqref{eq:dft_measurement} is revised as
\begin{equation}
    \mathbf{y} = \mathbf{\Theta} \mathbf{h}_{\mathrm{JAS}} + \mathbf{n},
\label{eq:measurement_jas}
\end{equation}
where $\mathbf{\Theta}= \mathbf{A}\mathbf{F}_{\mathrm{JAS}}\in \mathbb{C}^{M \times J}$ denotes the effective measurement matrix. A classical approach to solving \eqref{eq:measurement_jas} is OMP, which iteratively selects the codeword with the largest correlation to the residual signal. However, directly applying this serial correlation-based strategy to the JAS-domain representation is challenging. From \eqref{eq:same_subarray}, the first-null main-lobe width of the angular response within one subarray is $4/N_{\mathrm b}$, whereas that of the conventional full-array DFT codeword is $4/N$. Thus, the angular response of a JAS codeword is broadened by a factor of $N_{\mathrm{sub}}$, reflecting a localization-resolution tradeoff similar to that in STFT. To illustrate this effect on the correlation response used for OMP selection, Fig.~\ref{fig:codebook_comparison} depicts the magnitude distributions of $(\mathbf{A}\mathbf{F}_{\mathrm{DFT}})^{\mathrm{H}}\mathbf{y}$ and $(\mathbf{A}\mathbf{F}_{\mathrm{JAS}})^{\mathrm{H}}\mathbf{y}$, respectively. It can be observed that the response of each true path in $|(\mathbf{A}\mathbf{F}_{\mathrm{JAS}})^{\mathrm{H}}\mathbf{y}|$ spreads over more neighboring angle groups than that in the DFT-domain response, resulting in broader and less isolated peaks. Since OMP has no backtracking mechanism, such ambiguity may cause an early wrong selection that can bias the residual update and lead to error propagation. Therefore, serial greedy pursuit is not well suited for support detection over the JAS-domain representation.

To address the challenges in serial greedy pursuit, a natural alternative is to perform parallel sparse inference by jointly estimating all JAS coefficients. Since the JAS representation exhibits both angle-level sparsity and element-wise sparsity, the classical SGL formulation is considered~\cite{simon2013sparse}, which is given by
\begin{equation}
\begin{aligned}
    \min_{\mathbf{h}_{\mathrm{JAS}}} \, 
    &(\mathbf{y} - \mathbf{\Theta}\mathbf{h}_{\mathrm{JAS}})^{\mathrm{H}} 
    \mathbf{C}^{-1} 
    (\mathbf{y} - \mathbf{\Theta}\mathbf{h}_{\mathrm{JAS}})+\gamma_1 \|\mathbf{h}_{\mathrm{JAS}}\|_1 \\
    & + \gamma_2 \sum_{g=1}^G 
    \left\|[\mathbf{h}_{\mathrm{JAS}}]_{(g-1)N_{\mathrm{sub}}+1:gN_{\mathrm{sub}}}\right\|_2,
\end{aligned}
\label{eq:map_sgl}
\end{equation}
where $\gamma_1$ and $\gamma_2$ control the element-wise sparsity and the group-level angular sparsity, respectively. A typical way to solve \eqref{eq:map_sgl} is to use the iterative shrinkage-thresholding algorithm (ISTA)~\cite{zhang2018istanet}. Specifically, at the $k$-th iteration, the data fidelity term yields the following gradient update
\begin{equation}
\begin{aligned}
    \mathbf{r}^{(k)} &= \hat{\mathbf{h}}^{(k)}_{\mathrm{JAS}} + \eta \mathbf{\Theta}^{\mathrm{H}} \mathbf{C}^{-1} (\mathbf{y} - \mathbf{\Theta}\hat{\mathbf{h}}^{(k)}_{\mathrm{JAS}}) \\
    &= \hat{\mathbf{h}}^{(k)}_{\mathrm{JAS}} + \eta \mathbf{\Theta}^{\mathrm{H}}\mathbf{C}^{-1}\mathbf{\Theta} (\mathbf{h}_{\mathrm{JAS}} - \hat{\mathbf{h}}^{(k)}_{\mathrm{JAS}}) + \eta \mathbf{\Theta}^{\mathrm{H}} \mathbf{C}^{-1} \mathbf{n},
\end{aligned}
\label{eq:pgd_update}
\end{equation}
where $\eta>0$ is the step size and $\hat{\mathbf{h}}^{(k)}_{\mathrm{JAS}}$ is the estimated result of $\mathbf{h}_{\mathrm{JAS}}$ at the $k$-th iteration. Besides, $\mathbf{\Theta}^{\mathrm{H}}\mathbf{C}^{-1}\mathbf{\Theta}$ is the Gram matrix of the effective sensing dictionary. The updated intermediate vector $\mathbf{r}^{(k)}$ is then passed through a proximal mapping that combines element-wise and group-wise shrinkage to obtain the next estimate $\hat{\mathbf{h}}^{(k+1)}_{\mathrm{JAS}}$.

The limitation of this model-based parallel solver can be observed from \eqref{eq:pgd_update}. Due to the correlation among neighboring JAS grids, the term $\mathbf{\Theta}^{\mathrm{H}}\mathbf{C}^{-1}\mathbf{\Theta}(\mathbf{h}_{\mathrm{JAS}}-\hat{\mathbf{h}}^{(k)}_{\mathrm{JAS}})$ introduces inter-grid leakage into $\mathbf{r}^{(k)}$, such that true support components and leakage are entangled. Although the subsequent proximal mapping promotes sparsity, its fixed shrinkage rules cannot adapt to the varying path gains and leakage levels across grids and channel realizations. Consequently, strong leakage may be mistakenly retained as active supports, while weak true supports may be missed, leading to degraded precision and recall. This motivates us to develop an adaptive inference rule for reliable parallel support detection in the JAS domain.

\section{Proposed Two-Stage Framework}
\label{sec: learning}
This section presents the proposed signal processing and deep learning co-design framework for spatially non-stationary CE. We first provide an overview of the two-stage estimation framework, followed by the detailed implementations of its core modules. Finally, we introduce the training objective designed to handle the severe class imbalance caused by the sparsity of the high-frequency channel.

\subsection{Two-Stage Framework Overview}
\label{subsec:framework_overview}
\begin{figure*}[!t]
    \centering
    \includegraphics[width=1\textwidth]{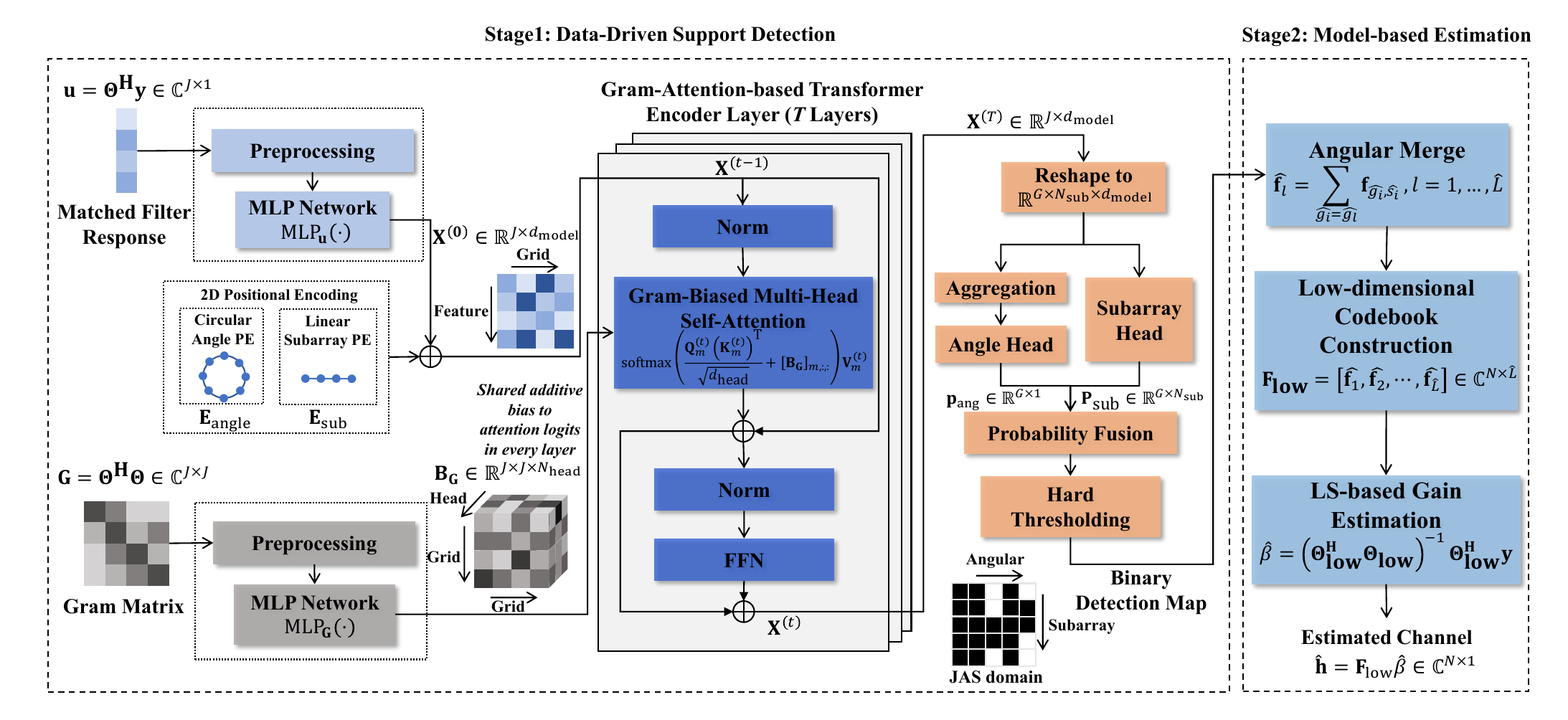} 
    \caption{The proposed two-stage CE framework, consisting of (left) data-driven parallel support detection and (right) model-based gain estimation.}
    \label{fig:diagram}
\end{figure*}

As depicted in Fig.~\ref{fig:diagram}, the proposed framework tackles spatially non-stationary CE by decomposing it into a two-stage process. In the first stage, a learning-based support detector evaluates all $J$ grids in parallel and outputs a binary support map
$\hat{\mathbf{B}}\in\{0,1\}^{G\times N_{\mathrm{sub}}}$, where $[\hat{\mathbf{B}}]_{g,s}=1$ indicates that the JAS codeword associated with the $(g,s)$-th angle-subarray pair is detected as active. Then, the corresponding support set is  obtained from the nonzero entries of $\hat{\mathbf{B}}$ as
\begin{equation}
    \mathcal{S}
    =
    \left\{(\hat{g}_k,\hat{s}_k)\mid
    [\hat{\mathbf{B}}]_{\hat{g}_k,\hat{s}_k}=1,\;
    k=1,2,\cdots,|\mathcal{S}|
    \right\},
\label{eq:support_set}
\end{equation}
where $|\mathcal{S}| \ll J$ due to the channel sparsity in the JAS domain.

In the second stage, the detected JAS codewords are first grouped according to their angle indices. Suppose that $\hat{L}$ distinct angle indices are detected from $\mathcal{S}$, and let $\hat{g}_l$ denote the $l$-th detected angle index. For each detected angle, the codewords associated with its active subarrays are summed to reconstruct the corresponding non-stationary steering vector as
\begin{equation}
    \hat{\mathbf{f}}_l = \sum_{\hat{g}_i = \hat{g}_l} \mathbf{f}_{\hat{g}_i, \hat{s}_i}, \quad l=1, 2, \dots, \hat{L},
\label{eq:path_reconstruction}
\end{equation}
where the summation is taken over all detected support pairs in $\mathcal{S}$ whose angle index equals $\hat{g}_l$. In this way, $\hat{\mathbf{f}}_l$ serves as the estimated result of $\mathbf{b}(\theta_l)\odot(\boldsymbol{\phi}_l\otimes\mathbf{1}_{N_{\mathrm b}})$ in \eqref{eq:vr-multi-path-channel}. By concatenating these non-stationary vectors, a low-dimensional codebook is obtained as
\begin{equation}
    \mathbf{F}_{\mathrm{low}} = \left[\hat{\mathbf{f}}_{1}, \hat{\mathbf{f}}_{2}, \cdots, \hat{\mathbf{f}}_{\hat{L}}\right] \in \mathbb{C}^{N \times \hat{L}},
\label{eq:ns_dictionary}
\end{equation}
based on which the sensing matrix is given by $\mathbf{\Theta}_{\mathrm{low}} = \mathbf{A}\mathbf{F}_{\mathrm{low}} \in \mathbb{C}^{M \times \hat{L}}$. In this regard, the number of unknown variables drastically reduces from $J$ to $\hat{L}$. Due to the sparsity of the high-frequency channel, the number of paths is typically restricted to a small integer~\cite{7959169}, which is much smaller than the number of measurement. Consequently, the complex path gains can be obtained via the LS estimator robustly, given by
\begin{equation}
    \hat{\boldsymbol{\beta}} = \left(\mathbf{\Theta}_{\mathrm{low}}^{\mathrm{H}} \mathbf{\Theta}_{\mathrm{low}}\right)^{-1} \mathbf{\Theta}_{\mathrm{low}}^{\mathrm{H}} \mathbf{y},
\label{eq:ls_estimation_first}
\end{equation}
where $\hat{\boldsymbol{\beta}} \in \mathbb{C}^{\hat{L}\times 1}$ contains the estimated complex gains associated with the $\hat{L}$ detected paths. Finally, the spatial channel is recovered as $\hat{\mathbf{h}} = \mathbf{F}_\mathrm{low} \hat{\boldsymbol{\beta}} \in \mathbb{C}^{N\times 1}$. Within this two-stage framework, the fundamental challenge lies in the first stage, as the second stage relies on a relatively deterministic signal processing procedure. Therefore, the remainder of this subsection focuses on the overview of the first stage.

We begin with specifying the inputs for the first stage by revisiting the data fidelity term in \eqref{eq:map_sgl}. By expanding the weighted quadratic form in \eqref{eq:map_sgl}, we obtain
\begin{equation}
\begin{aligned}
    &(\mathbf{y} - \mathbf{\Theta}\mathbf{h}_{\text{JAS}})^{\mathrm{H}} \mathbf{C}^{-1} (\mathbf{y} - \mathbf{\Theta}\mathbf{h}_{\text{JAS}}) \\
    &= \mathbf{y}^{\mathrm{H}} \mathbf{C}^{-1} \mathbf{y} - 2\Re\big\{\mathbf{h}_{\text{JAS}}^{\mathrm{H}} \mathbf{\Theta}^{\mathrm{H}} \mathbf{C}^{-1} \mathbf{y}\big\} + \mathbf{h}_{\text{JAS}}^{\mathrm{H}} \mathbf{\Theta}^{\mathrm{H}} \mathbf{C}^{-1} \mathbf{\Theta} \mathbf{h}_{\text{JAS}}.
\end{aligned}
\label{eq:map_expansion}
\end{equation}

It can be observed from \eqref{eq:map_expansion} that $\mathbf{h}_{\text{JAS}}$ is exclusively coupled with the observation $\mathbf{y}$ and the dictionary $\mathbf{\Theta}$ through two specific terms: The matched filter response $\mathbf{\Theta}^{\mathrm{H}} \mathbf{C}^{-1} \mathbf{y}$ and the Gram matrix $\mathbf{\Theta}^{\mathrm{H}} \mathbf{C}^{-1} \mathbf{\Theta}$. According to the Neyman-Fisher factorization theorem~\cite{kay1993estimation}, these two terms constitute the sufficient statistics for the support detection problem. Consequently, feeding these two terms into the network, rather than the raw high-dimensional triplet $(\mathbf{y}, \mathbf{A}, \mathbf{F}_{\text{JAS}})$, significantly compresses the input space without loss of statistical information. However, directly computing these statistics involves the inverse of the noise covariance matrix $\mathbf{C}$. To avoid this additional computational burden, we apply digital baseband whitening. For the $q$-th time slot, we decompose $\mathbf{W}_{\mathrm{RF},q}\mathbf{W}_{\mathrm{RF},q}^{\mathrm{H}}$ as
\begin{equation}
    \mathbf{W}_{\mathrm{RF},q}\mathbf{W}_{\mathrm{RF},q}^{\mathrm{H}}
    =
    \mathbf{U}_q\mathbf{\Sigma}_q\mathbf{U}_q^{\mathrm{H}},
\label{eq:svd_decomposition}
\end{equation}
where $\mathbf{U}_q$ is a unitary matrix and $\mathbf{\Sigma}_q$ is a diagonal matrix containing the nonnegative eigenvalues of $\mathbf{W}_{\mathrm{RF},q}\mathbf{W}_{\mathrm{RF},q}^{\mathrm{H}}$. Then, we set the digital combiner as $\mathbf{W}_{\mathrm{BB},q}=\mathbf{U}_q\mathbf{\Sigma}_q^{-1/2}\mathbf{U}_q^{\mathrm{H}}$. In this regard, the effective noise covariance becomes $\mathbf{C}=\sigma_{\mathrm n}^2\mathbf{I}_M$, and the network inputs simplifies to $\mathbf{u}=\mathbf{\Theta}^{\mathrm{H}}\mathbf{y}\in\mathbb{C}^{J\times 1}$ and $\mathbf{G}=\mathbf{\Theta}^{\mathrm{H}}\mathbf{\Theta}\in\mathbb{C}^{J\times J}$. Having specified $(\mathbf{u},\mathbf{G})$ as the network inputs, we next outline the three core design principles of the proposed learning architecture.

First, we lift the matched filter response $\mathbf{u}$ into a high-dimensional latent space. Although $\mathbf{u}$ provides an initial response over all JAS grids, each grid is represented by only one complex scalar, where the desired support information and leakage components are mixed together. Therefore, to improve feature separability, a point-wise embedding function $\mathcal{M}_{\text{embed}}(\cdot)$ maps $\mathbf{u}$ into a high-dimensional real-valued tensor as
\begin{equation}
    \mathbf{X}^{(0)} = \mathcal{M}_{\text{embed}}(\mathbf{u}) \in \mathbb{R}^{J \times d_{\text{model}}},
\label{eq:feature_mapping}
\end{equation}
where $\mathcal{M}_{\text{embed}}(\cdot)$ denotes the embedding function, and $d_{\text{model}}$ is the feature dimension. 

Second, we design a Gram-guided multi-head self-attention (MHSA) mechanism, termed Gram-attention, to refine the entangled grid features. The key motivation is the structural consistency between the physical leakage pattern and the attention operation. Specifically, the grid-to-grid leakage relationship is encoded in the Gram matrix $\mathbf{G}$, while the token-to-token interaction in MHSA is described by the attention map. This correspondence makes attention a natural mechanism for modeling interference among JAS grids. By injecting the Gram matrix as a physical prior into MHSA, the feature interactions are guided not only by learned feature similarity but also by the correlation structure of the effective sensing matrix. Based on the proposed Gram-attention, the feature update at the $t$-th layer is given by
\begin{equation}
    \mathbf{X}^{(t)} = \mathcal{H}^{(t)} \big(\mathbf{X}^{(t-1)}, \mathbf{G}\big), \quad t=1,2,\cdots,T,
\label{eq:gram_attention}
\end{equation}
where $\mathcal{H}^{(t)}(\cdot)$ denotes the mapping of the $t$-th backbone layer, which is implemented as a Transformer encoder layer with the proposed Gram-attention mechanism. Each layer refines the grid features by suppressing the leakage-induced components. After $T$ layers of successive refinement, the final tensor $\mathbf{X}^{(T)} \in \mathbb{R}^{J \times d_{\text{model}}}$ contains a purified and discriminative feature representation for each JAS grid.

Third, to align with the prior function in \eqref{eq:map_sgl}, we design a dual-head structure. Specifically, the first head enforces the group-level angular sparsity by aggregating $\mathbf{X}^{(T)}$ into an angle-level representation $\mathbf{X}_{\mathrm{ang}} \in \mathbb{R}^{G \times d_{\text{model}}}$, outputting a probability score for each sampled angle. Simultaneously, the second head enforces the element-wise sparsity without any feature aggregation, directly mapping $\mathbf{X}^{(T)}$ to a probability score for each angle-subarray pair. Finally, these two scores are fused to yield the overall support probability $\mathbf{P} \in [0, 1]^{G \times N_{\mathrm{sub}}}$. By applying the hard thresholding on $\mathbf{P}$, the discrete support set $\mathcal{S}$ formulated in \eqref{eq:support_set} can be determined readily. In the following, the detailed implementations of these three core components will be elaborated respectively.

\subsection{Input Embedding}
\label{sub:embedding}
The embedding module $\mathcal{M}_{\text{embed}}(\cdot)$ that transforms the complex matched filter response $\mathbf{u}$ into high-dimensional features consists of three steps: Complex-to-real transformation, point-wise multi-layer perceptron (MLP) projection, and 2D positional encoding (PE).

First, since standard neural networks operate on real-valued tensors, each complex response is converted into a four-dimensional (4D) real-valued representation as
\begin{equation}
    \mathcal{D}(u_i) = \left[ \Re\{u_i\}, \, \Im\{u_i\}, \, |u_i|, \, \angle u_i \right]^{\mathrm{T}} \in \mathbb{R}^{4 \times 1},
\label{eq:unified_operator}
\end{equation}
where $u_i \in \mathbb{C}$ denotes the $i$-th element of $\mathbf{u}$. Subsequently, this 4D feature vector is projected to a $d_{\text{model}}$-dimensional token via $\text{MLP}_{\mathbf{u}}(\cdot)$, given by
\begin{equation}
    \boldsymbol{\xi}_i = \text{MLP}_{\mathbf{u}}\left(\mathcal{D}(u_i)\right) \in \mathbb{R}^{d_{\text{model}} \times 1}.
\label{eq:mlp_projection}
\end{equation}

In this paper, $\text{MLP}_{\mathbf{u}}(\cdot)$ is implemented as a two-layer linear network, interleaved with a layer normalization (LayerNorm) module and a Gaussian error linear unit (GELU) activation function, which introduces essential non-linearity to enhance the representational capacity of the network. Finally, by leveraging parallel tensor computation, this point-wise mapping is executed simultaneously across all $J$ grids. The resulting tokens are stacked to form the raw high-dimensional feature tensor $\mathbf{X}_{\text{raw}} = [(\boldsymbol{\xi}_1)^{\mathrm{T}}, (\boldsymbol{\xi}_2)^{\mathrm{T}}, \cdots, (\boldsymbol{\xi}_J)^{\mathrm{T}}]^{\mathrm{T}} \in \mathbb{R}^{J \times d_{\text{model}}}$.

Since the subsequent backbone adopts a Transformer encoder, PE is introduced to inject positional information into the input features, as self-attention cannot distinguish the ordering of JAS grids by itself~\cite{vaswani2017attention}. Notably, in the JAS domain, the angular dimension is cyclic due to angular periodicity, whereas the subarray dimension follows the linear layout of the antenna array. Accordingly, we adopt a factorized 2D PE scheme to encode the angular and subarray structures, respectively.

First, to model the cyclic topology of the angular domain, a circular PE matrix $\mathbf{E}_{\text{ang}} \in \mathbb{R}^{G \times d_{\text{model}}}$ is constructed. For the $g$-th angle index, its PE matrix is formulated as
\begin{equation}
    \left\{
    \begin{aligned}
        \big[\mathbf{E}_{\text{ang}}\big]_{g, 2p} &= \sin\left(2\pi \frac{g}{G} (p+1)\right) \\
        \big[\mathbf{E}_{\text{ang}}\big]_{g, 2p+1} &= \cos\left(2\pi \frac{g}{G} (p+1)\right)
    \end{aligned}
    \right.,
\label{eq:pe_angle}
\end{equation}
where $p = 0, 1, \cdots, d_{\text{model}}/2-1$ denotes the frequency index. Second, for the subarray domain, we employ the standard sinusoidal PE to construct $\mathbf{E}_{\text{sub}} \in \mathbb{R}^{N_{\mathrm{sub}} \times d_{\text{model}}}$~\cite{vaswani2017attention}. For the $s$-th subarray index, the PE matrix is given by
\begin{equation}
    \left\{
    \begin{aligned}
        \big[\mathbf{E}_{\text{sub}}\big]_{s, 2p} &= \sin\left(\frac{s}{10000^{2p/d_{\text{model}}}}\right) \\
        \big[\mathbf{E}_{\text{sub}}\big]_{s, 2p+1} &= \cos\left(\frac{s}{10000^{2p/d_{\text{model}}}}\right)
    \end{aligned}
    \right..
\label{eq:pe_sub}
\end{equation}

Finally, the PE matrices are expanded according to the block-wise ordering of the JAS codebook in \eqref{eq:jas-codebook}. Specifically, $\bar{\mathbf{E}}_{\text{ang}}\in \mathbb{R}^{J \times d_{\text{model}}}$ is obtained by repeating each row of $\mathbf{E}_{\text{ang}}$ for $N_{\mathrm{sub}}$ times, while $\bar{\mathbf{E}}_{\text{sub}}\in \mathbb{R}^{J \times d_{\text{model}}}$ is obtained by tiling $\mathbf{E}_{\text{sub}}$ for $G$ times. The initial feature tensor fed into the backbone is then given by
\begin{equation}
    \mathbf{X}^{(0)} = \mathbf{X}_{\text{raw}} + \bar{\mathbf{E}}_{\text{ang}} + \bar{\mathbf{E}}_{\text{sub}} \in \mathbb{R}^{J \times d_{\text{model}}}.
\label{eq:pe_fusion}
\end{equation}

In this way, each JAS grid feature carries both its response information and its relative position in the JAS domain, providing structured input for the subsequent Gram-attention-based backbone.

\subsection{Transformer Encoder Layer with Gram-Attention}
\label{sub:purification}
\begin{figure*}[!t]
    \centering
    \includegraphics[width=0.81\textwidth]{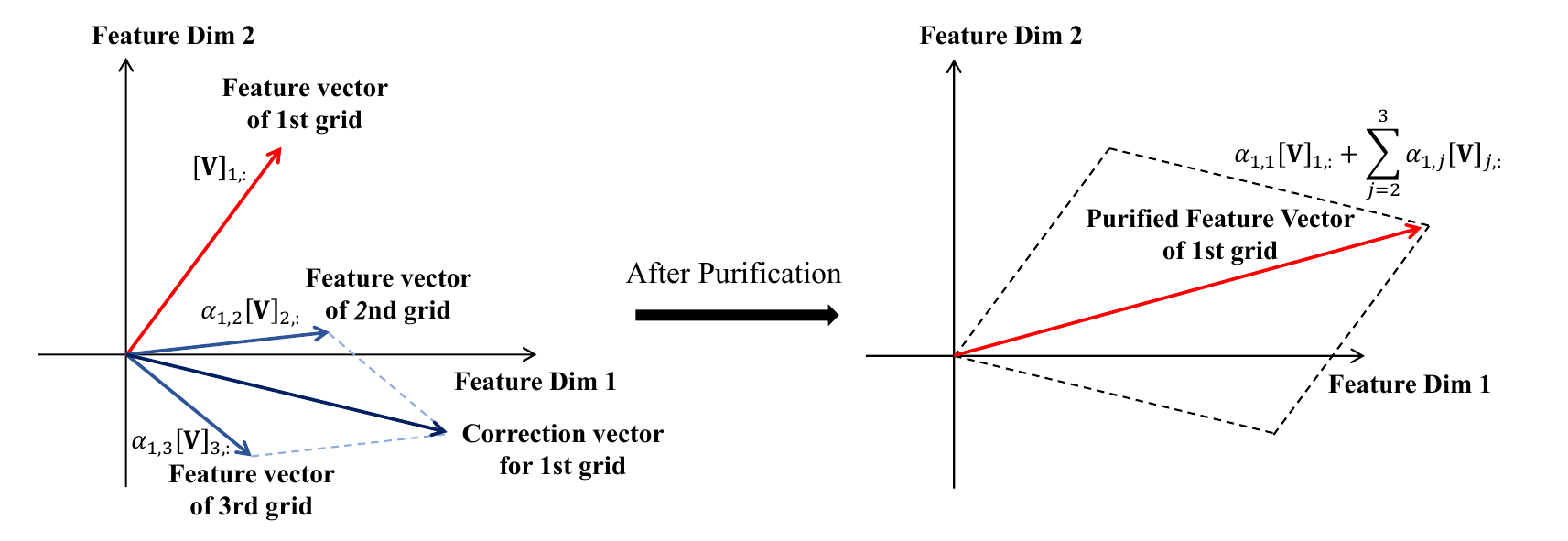} 
    \caption{Illustration for the procedure of leakage suppression, where only two dimensions and three grids are depicted here for the sake of simplicity. Besides, the indices for head and layer, i.e., $m$ and $t$, are omitted here for brevity.}
    \label{fig:feature_vector}
\end{figure*}

After the input embedding, each JAS grid is represented by a high-dimensional feature vector. However, the useful support evidence and leakage-related components are still mixed in these features. To purify the grid representations, we build the backbone as a stack of Transformer encoder layers, where each layer is equipped with the proposed Gram-attention mechanism. Suppose that the Gram-attention module contains $N_{\mathrm{head}}$ parallel heads, each operating in a subspace of dimension $d_{\mathrm{head}}=d_{\mathrm{model}}/N_{\mathrm{head}}$. For the $m$-th head at the $t$-th layer, the input features $\mathbf{X}^{(t-1)}$ are linearly projected into the query, key, and value matrices as
\begin{equation}
\begin{aligned}
    \mathbf{Q}_m^{(t)} &= \mathbf{X}^{(t-1)} \mathbf{W}_{m}^{Q,(t)}, \\
    \mathbf{K}_m^{(t)} &= \mathbf{X}^{(t-1)} \mathbf{W}_{m}^{K,(t)}, \\
    \mathbf{V}_m^{(t)} &= \mathbf{X}^{(t-1)} \mathbf{W}_{m}^{V,(t)}.
\end{aligned}
\label{eq:qkv_projection}
\end{equation}

Instead of relying solely on data-driven semantic correlations, we explicitly guide the attention mechanism using the deterministic physical leakage prior $\mathbf{G}$. Reusing the complex-to-real operator $\mathcal{D}(\cdot)$ in Section~\ref{sub:embedding}, each entry $[\mathbf{G}]_{i,j}$ is first transformed to a 4D descriptor $\mathcal{D}([\mathbf{G}]_{i,j})$. Subsequently, a small MLP network projects this descriptor into the bias vector as $[\mathbf{B}_{\mathbf{G}}]_{:, i, j} = \text{MLP}_{\mathbf{G}}(\mathcal{D}([\mathbf{G}]_{i,j})) \in \mathbb{R}^{N_{\mathrm{head}} \times 1}$. In this work, $\text{MLP}_{\mathbf{G}}(\cdot)$ is implemented as a two-layer MLP with a GELU activation function. Consequently, the Gram matrix is mapped into a multi-head bias tensor $\mathbf{B}_{\mathbf{G}} \in \mathbb{R}^{N_{\mathrm{head}} \times J \times J}$ through parallel computation over all grids. By introducing such physical bias tensor into the standard MHSA mechanism, the output $\mathbf{H}_m^{(t)} \in \mathbb{R}^{J \times d_{\mathrm{head}}}$ of the $m$-th head is obtained as
\begin{equation}
    \mathbf{H}_m^{(t)}
    =
    \mathrm{softmax}\left(
    \frac{\mathbf{Q}_m^{(t)}(\mathbf{K}_m^{(t)})^{\mathrm{T}}}{\sqrt{d_{\mathrm{head}}}}
    +
    [\mathbf{B}_{\mathbf{G}}]_{m,:,:}
    \right)
    \mathbf{V}_m^{(t)}.
\label{eq:head_output}
\end{equation}

To further examine how Gram-attention updates each grid feature, let $\alpha_{i,j}^{(t,m)}$ denote the $(i,j)$-th entry of the attention matrix in \eqref{eq:head_output}. For the $i$-th grid, the corresponding output feature can be decomposed as
\begin{equation}
    [\mathbf{H}_m^{(t)}]_{i,:}
    =
    \alpha_{i,i}^{(t,m)}[\mathbf{V}_m^{(t)}]_{i,:}
    +
    \sum_{j=1,\,j\neq i}^{J}
    \alpha_{i,j}^{(t,m)}
    [\mathbf{V}_m^{(t)}]_{j,:}.
\label{eq:vector_cancellation_decomposed}
\end{equation}

As shown in Fig.~\ref{fig:feature_vector}, the first term corresponds to the feature vector of the $i$-th grid, while the second term forms a learned correction term contributed by the other grids. Since both the attention weights and value projections are learnable, this correction term adaptively adjusts the representation of the $i$-th grid in the feature space, so that leakage-related components can be weakened and support-relevant evidence can be better preserved.

Then, the outputs of all attention heads at the $t$-th layer are concatenated and linearly projected as
\begin{equation}
    \mathbf{O}^{(t)}
    =
    \mathrm{Concat}\left(
    \mathbf{H}_1^{(t)},\mathbf{H}_2^{(t)},\cdots,\mathbf{H}_{N_{\mathrm{head}}}^{(t)}
    \right)
    \mathbf{W}^{O,(t)},
\label{eq:multi_head_concat}
\end{equation}
where $\mathbf{W}^{O,(t)}\in\mathbb{R}^{d_{\mathrm{model}}\times d_{\mathrm{model}}}$ is the output projection matrix of the $t$-th layer. Let $\mathcal{A}_{\mathrm{Gram}}^{(t)}(\cdot)$ denote the Gram-attention operator defined by \eqref{eq:qkv_projection}, \eqref{eq:head_output}, and \eqref{eq:multi_head_concat}. Based on this operator, the layer mapping $\mathcal{H}^{(t)}(\cdot)$ in \eqref{eq:gram_attention} is implemented by integrating Gram-attention, residual connections, LayerNorm, and a feed-forward network (FFN), given by
\begin{equation}
\begin{aligned}
    \mathbf{X}^{(t)\prime}
    &=
    \mathbf{X}^{(t-1)}
    +
    \mathcal{A}_{\mathrm{Gram}}^{(t)}
    \left(\mathrm{LN}(\mathbf{X}^{(t-1)}),\mathbf{B}_{\mathbf{G}}\right), \\
    \mathbf{X}^{(t)}
    &=
    \mathbf{X}^{(t)\prime}
    +
    \mathrm{FFN}^{(t)}
    \left(\mathrm{LN}(\mathbf{X}^{(t)\prime})\right),
\end{aligned}
\label{eq:transformer_layer}
\end{equation}
where $\mathrm{LN}(\cdot)$ denotes the LayerNorm operator and $\mathrm{FFN}^{(t)}(\cdot)$ is implemented as a two-layer MLP, where the feature dimension is first expanded from $d_{\mathrm{model}}$ to $4d_{\mathrm{model}}$, activated by GELU, and then projected back to $d_{\mathrm{model}}$. By cascading $T$ layers, the overall feature purification process is expressed as
\begin{equation}
    \mathbf{X}^{(T)}
    =
    \left(
    \mathcal{H}^{(T)}
    \circ
    \mathcal{H}^{(T-1)}
    \circ
    \cdots
    \circ
    \mathcal{H}^{(1)}
    \right)
    \left(\mathbf{X}^{(0)},\mathbf{B}_{\mathbf{G}}\right).
\label{eq:nested_operator}
\end{equation}

The final tensor $\mathbf{X}^{(T)}\in\mathbb{R}^{J\times d_{\mathrm{model}}}$ contains the purified feature representation of each JAS grid and is utilized for the subsequent support detection.

\subsection{Decoupled Heads and Loss Functions}
\label{sub:dual_head_loss}
To streamline the formulations in this subsection, we first reshape the 2D tensor $\mathbf{X}^{(T)}$ into a 3D tensor $\tilde{\mathbf{X}} \in \mathbb{R}^{G \times N_{\mathrm{sub}} \times d_{\text{model}}}$, where $[\tilde{\mathbf{X}}]_{g,s,:} \in \mathbb{R}^{d_{\text{model}} \times 1}$ denotes the refined feature vector for the $(g, s)$-th angle-subarray pair. Subsequently, a decoupled head structure is designed to enforce the group-level and element-wise sparsity formulated in \eqref{eq:map_sgl}.

\subsubsection{Angle Head for Group-Level Sparsity} 
The first head evaluates the existence of a path at the $g$-th sampled angle. Since a non-stationary path may only illuminate some subarrays, standard mean pooling is inadequate as averaging the valid features with the remaining empty ones may dilute the effective information. Thus, we employ a dynamic attention pooling (AP) mechanism. Specifically, a scoring network $\text{MLP}_{\text{attn}}(\cdot)$ followed by a softmax function evaluates the feature $[\tilde{\mathbf{X}}]_{g,s,:}$ to compute the normalized attention coefficients $w_{g,s}$ where $\sum_{s=1}^{N_{\mathrm{sub}}} w_{g,s}=1$. Then, the features $\tilde{\mathbf{X}}$ are aggregated into the angle-level representation $\mathbf{X}_{\text{ang}}$ as
\begin{equation}
    [\mathbf{X}_{\text{ang}}]_{g,:} = \sum_{s=1}^{N_{\mathrm{sub}}} w_{g,s} [\tilde{\mathbf{X}}]_{g,s,:},
\label{eq:angle_aggregation}
\end{equation}
which is then mapped to the group-level probability $\mathbf{p}_{\text{ang}} \in [0,1]^{G \times 1}$ via a small MLP followed by the sigmoid activation
\begin{equation}
    [\mathbf{p}_{\text{ang}}]_g = \mathrm{sigmoid} \left( \text{MLP}_{\text{ang}}([\mathbf{X}_{\text{ang}}]_g) \right).
\label{eq:p_angle}
\end{equation}

\subsubsection{Subarray Head for Element-Wise Sparsity} 
The second head evaluates the visibility of the $g$-th angle at the $s$-th subarray. Unlike the angle head which requires aggregation of the features before detection, we use another MLP network $\text{MLP}_{\text{sub}}(\cdot)$ to map the feature tensor $\tilde{\mathbf{X}}$ to the 2D probability map $\mathbf{P}_{\text{sub}}  \in [0,1]^{G \times N_{\mathrm{sub}}}$ directly, which is given by
\begin{equation}
    [\mathbf{P}_{\text{sub}}]_{g,s} = \mathrm{sigmoid}\left( \text{MLP}_{\text{sub}}([\tilde{\mathbf{X}}]_{g,s,:}) \right).
\label{eq:p_sub}
\end{equation}

In this work, $\mathrm{MLP}_{\mathrm{attn}}(\cdot)$, $\mathrm{MLP}_{\mathrm{ang}}(\cdot)$, and $\mathrm{MLP}_{\mathrm{sub}}(\cdot)$ are all composed of two linear layers with an intermediate GELU activation.

\subsubsection{Loss Functions and Probability Fusion}
During the training phase, the model is optimized via decoupled loss functions. Let $\mathbf{T}_{\mathrm{JAS}}\in\{0,1\}^{G\times N_{\mathrm{sub}}}$ denote the 2D ground-truth support map in the JAS domain. Since an angle grid is active as long as at least one of its subarrays is visible, the group-level label $\mathbf{t}_{\mathrm{ang}}\in\{0,1\}^{G\times 1}$ is derived as
\begin{equation}
    [\mathbf{t}_{\mathrm{ang}}]_g
    =
    \mathbb{I}\left(
    \sum_{s=1}^{N_{\mathrm{sub}}}
    [\mathbf{T}_{\mathrm{JAS}}]_{g,s}
    >0
    \right),
\label{eq:angle_label}
\end{equation}
where $\mathbb{I}(\cdot)$ denotes the indicator function. To address the severe positive-negative class imbalance caused by the extreme sparsity of the high-frequency channel, we employ the asymmetric loss (ASL) for angle-level support detection~\cite{9710171}, given by
\begin{equation}
\label{eq:asl_loss}
\begin{aligned}
    \mathcal{L}_{\mathrm{ang}}
    =
    &-\frac{1}{G}\sum_{g=1}^{G}
    \Big[
    [\mathbf{t}_{\mathrm{ang}}]_g
    \left(1-[\mathbf{p}_{\mathrm{ang}}]_g\right)^{\gamma^+}
    \log\left([\mathbf{p}_{\mathrm{ang}}]_g\right) \\
    &+
    \left(1-[\mathbf{t}_{\mathrm{ang}}]_g\right)
    \left([\mathbf{p}_{\mathrm{ang}}]_g^*\right)^{\gamma^-}
    \log\left(1-[\mathbf{p}_{\mathrm{ang}}]_g^*\right)
    \Big],
\end{aligned}
\end{equation}
where $[\mathbf{p}_{\mathrm{ang}}]_g^*=\max([\mathbf{p}_{\mathrm{ang}}]_g-\delta,0)$ introduces a probability margin $\delta$ to discard easily classified inactive angles. Moreover, $\gamma^+$ and $\gamma^-$ denote the focusing parameters for positive and negative samples, respectively. 

For subarray-level visibility prediction, a standard binary cross-entropy (BCE) loss over all angle-subarray grids is not appropriate, because subarray visibility is only defined when the corresponding angle is active. Therefore, we use a masked BCE loss, where only the subarray labels belonging to active angle groups are included in the loss calculation~\cite{7780460}, which is given by
\begin{equation}
\label{eq:masked_bce}
\begin{aligned}
    \mathcal{L}_{\mathrm{sub}}
    =
    -\frac{1}{N^{\prime}}
    &\sum_{g=1}^{G}
    \sum_{s=1}^{N_{\mathrm{sub}}}
    [\mathbf{t}_{\mathrm{ang}}]_g
    \Big[
    [\mathbf{T}_{\mathrm{JAS}}]_{g,s}
    \log\left([\mathbf{P}_{\mathrm{sub}}]_{g,s}\right) \\
    &+
    \left(1-[\mathbf{T}_{\mathrm{JAS}}]_{g,s}\right)
    \log\left(1-[\mathbf{P}_{\mathrm{sub}}]_{g,s}\right)
    \Big],
\end{aligned}
\end{equation}
where $N^{\prime}=N_{\mathrm{sub}}\sum_{g=1}^{G}[\mathbf{t}_{\mathrm{ang}}]_g$ is the number of grids involved in the masked BCE loss. Then, the overall training loss is given by
\begin{equation}
    \mathcal{L}
    =
    \kappa_{\mathrm{ang}}\mathcal{L}_{\mathrm{ang}}
    +
    \kappa_{\mathrm{sub}}\mathcal{L}_{\mathrm{sub}},
\label{eq:overall_loss}
\end{equation}
where $\kappa_{\mathrm{ang}}$ and $\kappa_{\mathrm{sub}}$ are balancing weights.

During inference, the two outputs are fused according to their decoupled training designs. Since the subarray head is supervised only within active angle groups, $[\mathbf{P}_{\mathrm{sub}}]_{g,s}$ naturally represents the conditional visibility probability of the $s$-th subarray given the existence of the $g$-th angle. Therefore, the final support probability of the $(g,s)$-th JAS grid is obtained by multiplying the angle-level probability with the conditional subarray-level probability, i.e.,
\begin{equation}
    \mathbf{P}
    =
    \mathrm{diag}(\mathbf{p}_{\mathrm{ang}})
    \mathbf{P}_{\mathrm{sub}}.
\label{eq:fusion_prob}
\end{equation}

Finally, applying a global threshold $\tau$ to $\mathbf{P}$ yields the binary support map $\hat{\mathbf{B}}$, from which the support set $\mathcal{S}$ in \eqref{eq:support_set} is obtained for the gain estimation in the second stage.

\section{Experimental Results}
\label{sec: simulations}
This section first presents the simulation setup and introduces the baselines for comparison. Then, simulation results and ablation studies are provided to evaluate the effectiveness of the proposed framework.

\subsection{Simulation Setup and Baseline Definitions}
\subsubsection{Simulation Setup}
We consider a communication system operating at a carrier frequency of $30$ GHz. The BS is equipped with a $128$-antenna ULA and $N_{\mathrm{RF}} = 8$ RF chains. The array is divided into $N_{\mathrm{sub}} = 8$ subarrays to characterize the spatial non-stationarity. Besides, the angular domain is quantized into $G = 128$ grids, resulting in a JAS codebook size of $J = 1024$. For each channel sample, the number of paths is a random integer uniformly distributed between $1$ and $5$. For each individual path, the spatial angle $\theta_l$ is uniformly generated within $[-1, 1]$, and the complex channel gain $g_l$ is drawn from $\mathcal{CN}(0, 1)$. To simulate the spatial non-stationarity, each path is assigned a random subarray visibility pattern, with both the number and positions of visible subarrays randomly generated. The number of pilot measurements $Q$ is uniformly sampled from $[8, 16]$. Additionally, the transmit power is set to $30$ dBm and the SNR of the signal is uniformly sampled between $0$ dB and $20$ dB. 

To prevent overfitting and enhance the generalization capability, the training data is generated via a dynamic streaming approach, where $20,000$ channel realizations are instantiated on-the-fly per epoch. For a fair and reproducible evaluation, fixed validation and test sets, each containing $10,000$ offline-generated samples, are utilized. The proposed model is configured with $T=6$ layers, $N_{\mathrm{head}}=8$ attention heads and a hidden dimension of $d_{\text{model}} = 128$. Besides, the model is trained end-to-end utilizing the AdamW optimizer with a batch size of $32$ and an initial learning rate of $3 \times 10^{-4}$. The training spans $200$ epochs, regulated by a cosine annealing learning rate scheduler with a $5$-epoch warmup. To stabilize the training process, the gradient norm is clipped at $2.0$. For the ASL in \eqref{eq:asl_loss}, the parameters are set to $\gamma^- = 2.0$ and $\gamma^+ = 0.0$ with a probability margin of $\delta = 0.05$, while the balancing weights in \eqref{eq:overall_loss} are fixed at $\kappa_{\text{ang}} = 1.0$ and $\kappa_{\text{sub}} = 1.0$, respectively. Furthermore, the hard threshold $\tau$ is searched on the validation set by maximizing the F1-score over $[0.05,0.95]$ with a step size of $0.01$, and is then fixed for the performance evaluations on the test set.

\subsubsection{Benchmark Definitions}
To evaluate the proposed framework\footnote{To facilitate reproducibility, the source code will be released upon acceptance of this paper.}, we compare it with eight representative baselines as follows\footnote{For a fair comparison, all considered baselines are implemented under the same two-stage protocol as the proposed method, namely support detection followed by LS-based gain estimation. Besides, Deep-SGL-ISTA, Deep-CFAR, and U-Net are trained with the same ground-truth support labels and the same ASL loss in \eqref{eq:asl_loss}.}:
\begin{itemize}
    \item \textbf{OMP-DFT}: Employing the classical OMP algorithm with the standard DFT codebook, which is utilized to illustrate the model mismatch induced by spatial non-stationarity.
    \item \textbf{OMP-JAS}: This baseline applies OMP to the JAS codebook. It serves to evaluate the limitation of serial greedy pursuit for support detection over the visibility-aware JAS representation.
    \item \textbf{SGL-ISTA}: Solving the SGL formulation in \eqref{eq:map_sgl} using ISTA. Each iteration consists of the gradient update in \eqref{eq:pgd_update}, followed by element-wise soft-thresholding and group-wise shrinkage.
    \item \textbf{Deep-SGL-ISTA}: Implements a data-driven variant of SGL-ISTA. It parameterizes the fixed update operator in SGL-ISTA by a trainable linear layer and replaces the proximal thresholds with learnable parameters~\cite{gregor2010learning,zhang2018istanet}, thereby improving adaptivity while retaining the SGL structure.
    \item \textbf{CA-CFAR}: Implementing the classical cell-averaging constant false alarm rate (CA-CFAR) detector~\cite{4102829}. For each JAS grid, CA-CFAR estimates a local threshold from neighboring reference cells after excluding guard cells. The support is then determined by comparing the grid response with the threshold.
    \item \textbf{Deep-CFAR}: Realizing a CFAR-inspired local CNN detector~\cite{8629967}, which utilizes a lightweight convolutional neural network (CNN) to predict the support probability of the grid from its surrounding local context.
    \item \textbf{U-Net}: Achieving a generic image-to-mask learning baseline based on the U-Net architecture~\cite{ronneberger2015u,10640941}. The transformed matched filter response $\mathcal{D}(\mathbf{u})$ is reshaped into a $\mathbb{R}^{4\times G\times N_{\mathrm{sub}}}$ tensor, which is then processed by a convolutional encoder-decoder with skip connections to predict the binary support map.
    \item \textbf{Oracle LS}: Performing the LS estimation with the ground-truth support set, serving as an upper bound for channel reconstruction performance.
\end{itemize}

\subsection{Support Detection Performance}
We first evaluate the accuracy of active support identification on the test set. All algorithms except Oracle LS are compared in terms of precision, recall, and F1-score~\cite{powers2020evaluation}. In addition, the average runtime of each algorithm is reported to provide a comparison of computational efficiency. The results are summarized in Table~\ref{tab:support_detection}.

As shown in Table~\ref{tab:support_detection}, the proposed method achieves the highest precision, recall, and F1-score among all baselines. OMP+DFT obtains relatively high recall but suffers from low precision due to the stationary DFT model mismatch, while OMP+JAS performs worse because the serial greedy selection cannot effectively handle the JAS-domain support detection problem. Compared with CFAR-based methods, the proposed method improves both precision and recall by a large margin, showing the advantage of global feature interaction over local thresholding. For sparse optimization-based methods, SGL-ISTA improves the precision compared with greedy pursuit, but its F1-score remains much lower than that of the proposed method. Moreover, its iterative shrinkage updates result in a runtime of $29.04$ ms, which is more than one order of magnitude slower than the proposed method. Although Deep-SGL-ISTA reduces the inference time through unfolding, its sparse-group recovery structure still limits the detection performance. Moreover, U-Net provides a competitive learning-based baseline, but it essentially treats support detection as an image-to-mask problem and does not explicitly exploit the prior embedded in the Gram matrix, leading to a clear performance gap from the proposed method. By using the Gram-domain prior to guide feature interaction, the proposed method achieves the best precision-recall tradeoff while maintaining millisecond-level inference.

\begin{table}[t]
\caption{Support Detection Performance on the Test Set and Runtime Comparison.}
\label{tab:support_detection}
\centering
\renewcommand{\arraystretch}{1.2}
\begin{tabular}{@{}l c c c c@{}}
\toprule
\textbf{Algorithm} & \textbf{Prec. (\%)} & \textbf{Rec. (\%)} & \textbf{F1 (\%)} & \textbf{Runtime (ms)} \\
\midrule
OMP+DFT          & 21.91 & 78.31 & 34.24 & 0.45 \\
OMP+JAS          & 13.26 & 29.62 & 18.32 & 6.45 \\
SGL-ISTA         & 35.18 & 53.01 & 42.29 & 29.04 \\
CA-CFAR          & 4.00  & 15.00 & 6.32  & 0.68 \\
Deep-SGL-ISTA    & 69.00 & 54.05 & 60.62 & 1.12 \\
Deep-CFAR        & 65.19 & 49.19 & 56.07 & \textbf{0.42} \\
U-Net            & 79.03 & 63.94 & 70.69 & 0.96 \\
\textbf{Proposed}& \textbf{91.29} & \textbf{81.89} & \textbf{86.34} & 2.14 \\
\bottomrule
\end{tabular}
\end{table}

\subsection{CE Performance versus SNR}
\begin{figure}[!t]
    \centering
    \includegraphics[height=0.65\columnwidth]{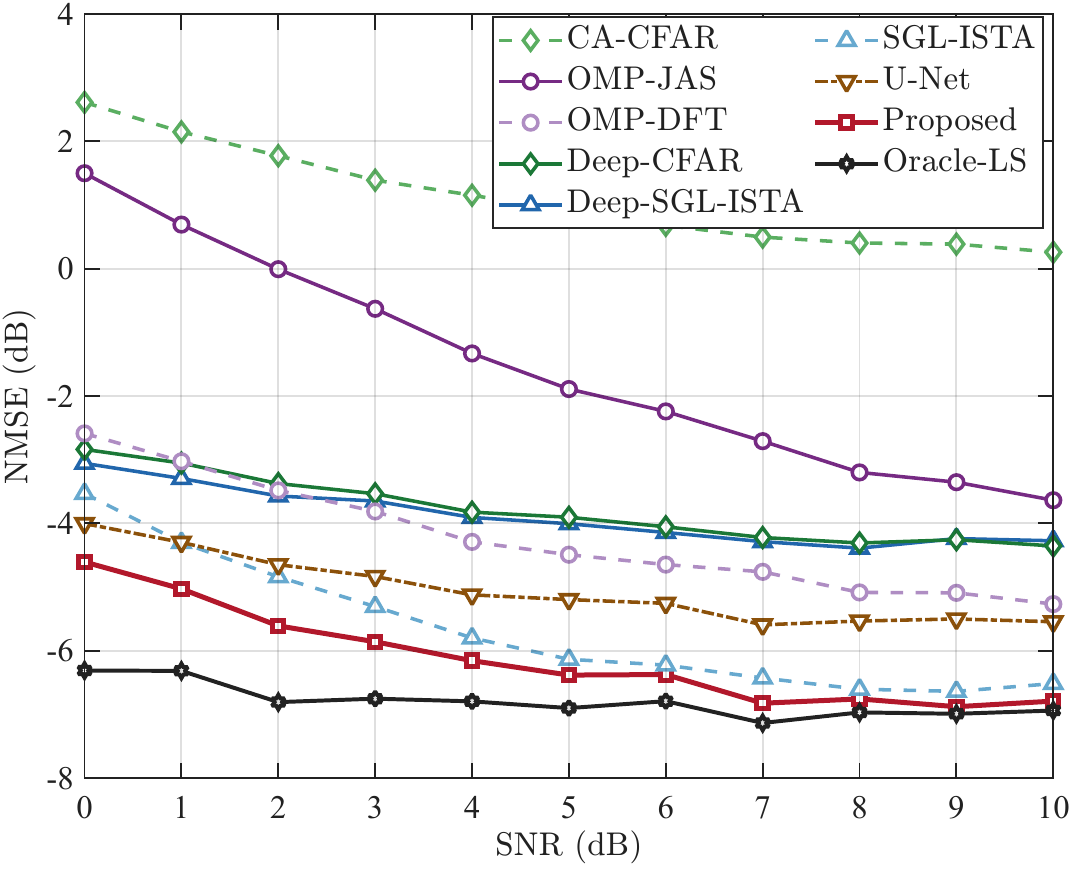} 
    \caption{CE NMSE versus SNR with fixed numbers of paths and pilot measurements, where the other parameters are set as $L=3$ and $Q=12$, respectively.}
    \label{fig:nmse_snr_fixed_LQ}
\end{figure}

\begin{figure}[!t]
    \centering
    \includegraphics[height=0.65\columnwidth]{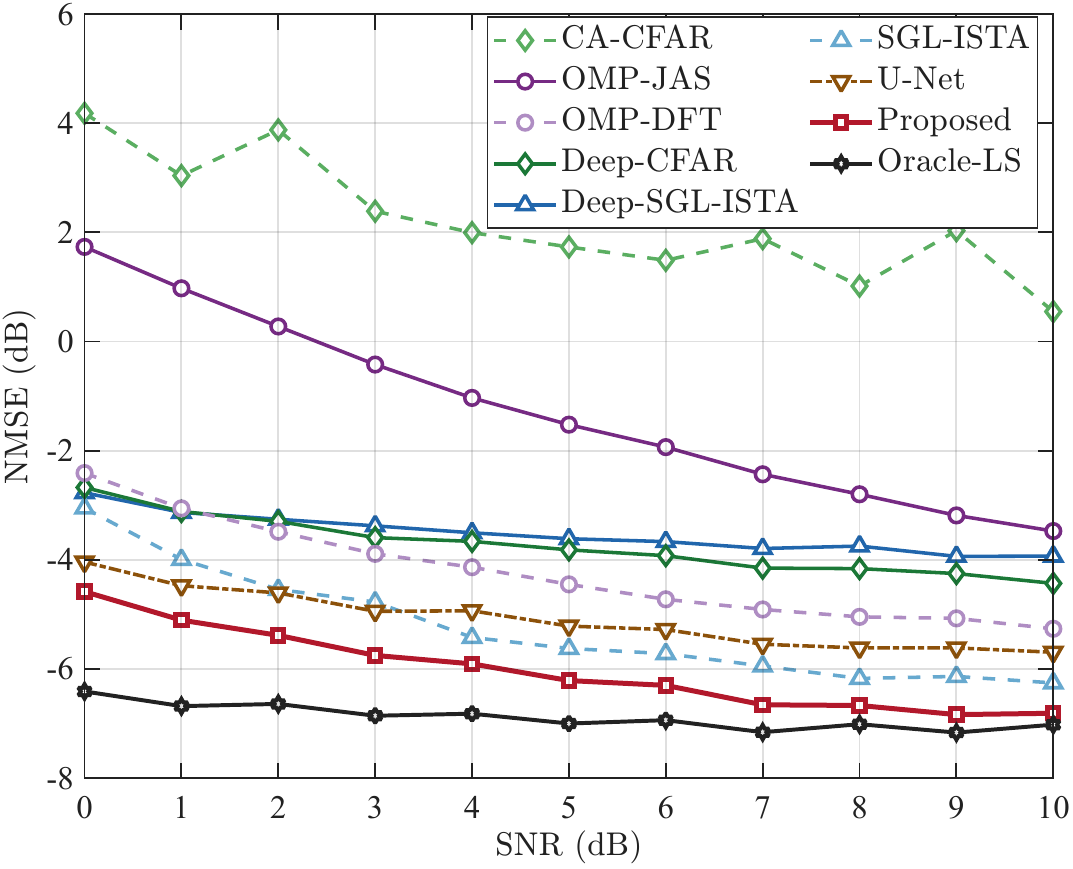} 
    \caption{CE NMSE versus SNR with random numbers of paths and pilot measurements, where $L$ and $Q$ are randomly generated for each channel realization.}
    \label{fig:nmse_snr_random_LQ}
\end{figure}

Having evaluated the accuracy of support detection, we subsequently compare the channel reconstruction accuracy using the normalized mean square error (NMSE), which is defined as
\begin{equation}
    \text{NMSE} = \frac{1}{N_{\text{mc}}} \sum_{i=1}^{N_{\text{mc}}} \frac{\|\mathbf{h}^{(i)} - \hat{\mathbf{h}}^{(i)}\|_2^2}{\|\mathbf{h}^{(i)}\|_2^2},
\label{eq:nmse}
\end{equation}
where $N_{\text{mc}}=1000$ represents the number of monte carlo trials, while $\mathbf{h}^{(i)}$ and $\hat{\mathbf{h}}^{(i)}$ denote the true and the estimated spatial channels for the $i$-th sample, respectively. Based on this metric, we investigate the algorithmic robustness across varying SNR levels, which are depicted in Fig.~\ref{fig:nmse_snr_fixed_LQ} and Fig.~\ref{fig:nmse_snr_random_LQ} under fixed and randomized parameter configurations, respectively.

As illustrated in Fig.~\ref{fig:nmse_snr_fixed_LQ}, we fix the number of paths to $L=3$ and the pilot measurements to $Q=12$, while scaling the SNR from $0$ dB to $10$ dB. It can be observed that the NMSE of all evaluated algorithms consistently decreases as the SNR increases. Crucially, the proposed method significantly outperforms all benchmarks across the entire SNR regime, and its performance is the closest to the Oracle LS bound. This suggests that the detected supports are sufficiently accurate to enable effective LS-based channel reconstruction. Among the benchmarks, U-Net and SGL-ISTA exhibit relatively competitive performance. Nevertheless, U-Net lacks explicit Gram-domain physical guidance, while SGL-ISTA is limited by fixed shrinkage rules, making both methods less effective than the proposed method. Besides, Deep-CFAR and Deep-SGL-ISTA suffer from an early performance plateau as the SNR increases, indicating their limited adaptability to non-stationary channel features. Traditional schemes such as OMP-DFT, OMP-JAS, and CA-CFAR are also less competitive, since their support decisions mainly rely on predefined matching or local statistics, rather than adaptive support inference over the JAS-domain representation.

To further validate the robustness of our architecture under dynamical environments, Fig.~\ref{fig:nmse_snr_random_LQ} evaluates the NMSE performance where $L$ and $Q$ are randomly generated for each channel realization. Despite the increased structural uncertainty, the proposed method still preserves its performance improvement over all baseline algorithms and tightly tracks the Oracle LS curve. This superior generalization capability indicates that the proposed Gram-attention mechanism effectively suppresses the dictionary-induced leakage and adapts to arbitrary combinations of path counts and pilot overheads without requiring model retraining.

\subsection{CE Performance versus Number of Pilots}
\begin{figure}[!t]
    \centering
    \includegraphics[height=0.65\columnwidth]{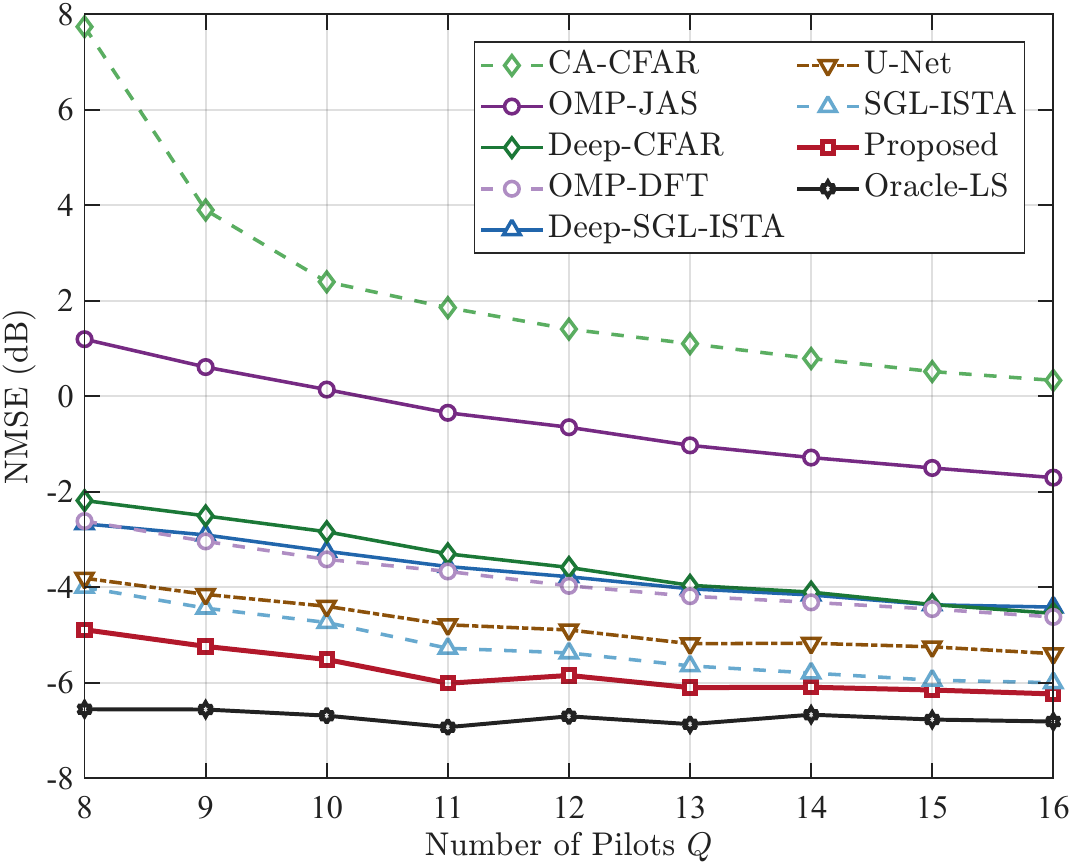} 
    \caption{CE NMSE versus the number of pilot measurements with fixed SNR and number of paths, where the other parameters are set as $\mathrm{SNR}=3$ dB and $L=3$, respectively.}
    \label{fig:nmse_Q_fixed_snr_L}
    
    \vspace{0.3cm} 

    \includegraphics[height=0.65\columnwidth]{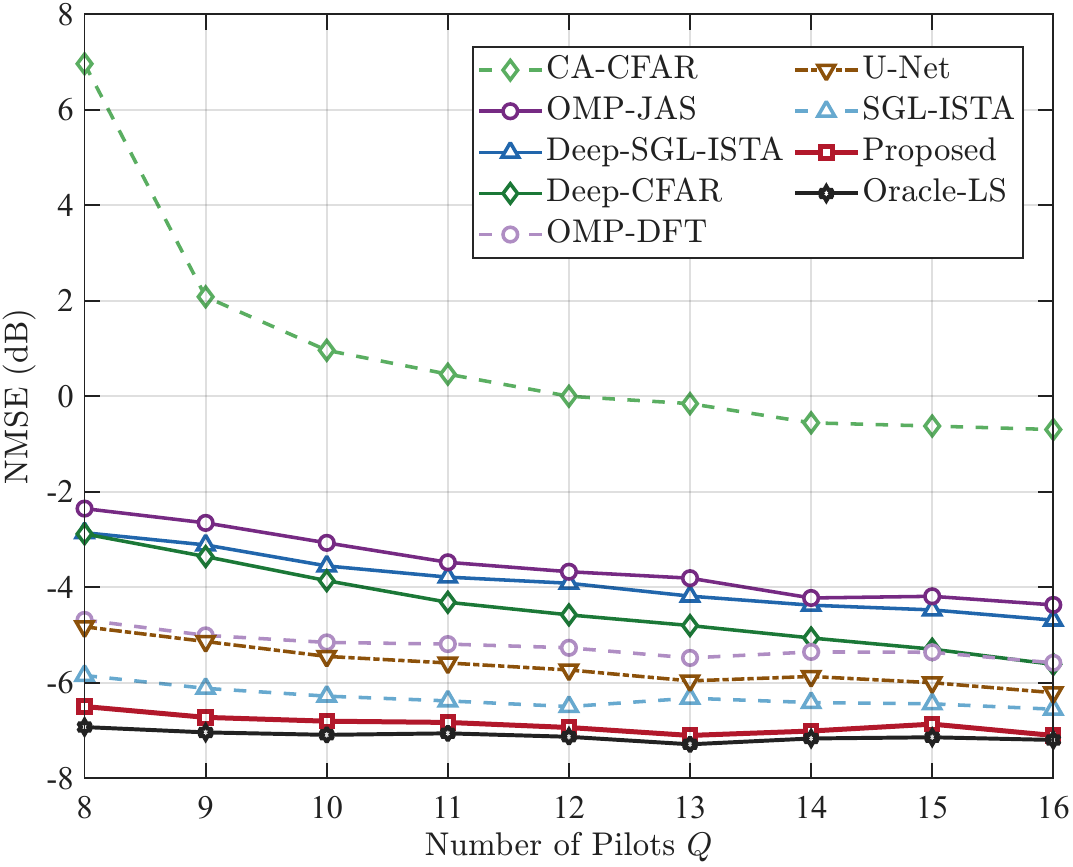} 
    \caption{CE NMSE versus the number of pilot measurements under random channel conditions, where the SNR and the number of paths are randomly generated for each channel realization.}
    \label{fig:nmse_Q_random_snr_L}
\end{figure}

To demonstrate the robustness of the proposed model under different number of measurements, we evaluate the CE performance across varying numbers of pilot $Q$. The simulation curves are presented in Fig.~\ref{fig:nmse_Q_fixed_snr_L} and Fig.~\ref{fig:nmse_Q_random_snr_L}, respectively.

Fig.~\ref{fig:nmse_Q_fixed_snr_L} shows the CE NMSE versus the number of pilot measurements $Q$ under the fixed setting of $\mathrm{SNR}=3$ dB and $L=3$. As $Q$ increases from $8$ to $16$, all algorithms exhibit improved NMSE performance, since more pilot measurements provide richer observations for support detection and gain estimation. The proposed method achieves the best performance over the whole range of $Q$. In particular, its NMSE decreases noticeably when $Q$ increases from $8$ to $11$, and then enters a relatively stable region, indicating that the proposed Gram-attention detector can obtain reliable support information with a moderate number of pilot measurements. In comparison, U-Net and SGL-ISTA also benefit from increasing $Q$, but their NMSE curves remain clearly separated from the proposed method. Additionally, OMP-DFT, Deep-SGL-ISTA, and Deep-CFAR also improve clearly with more pilot measurements, but their NMSE still saturates above the proposed method, indicating that the additional observations are not fully converted into performance improvement. Besides, OMP-JAS and CA-CFAR remain less competitive because their decisions rely on serial selection or local statistics.

Fig.~\ref{fig:nmse_Q_random_snr_L} further evaluates the NMSE performance under randomized channel conditions, where the SNR and the number of paths vary across channel realizations. Different from the fixed setting, the proposed method remains highly stable across the entire range of $Q$ and consistently stays close to the Oracle LS benchmark. This demonstrates that the proposed framework is robust to varying channel sparsity and noise conditions, rather than being effective only under a fixed channel configuration.

\subsection{Ablation Studies}

\begin{figure}[!t]
    \centering
    \includegraphics[height=0.65\columnwidth]{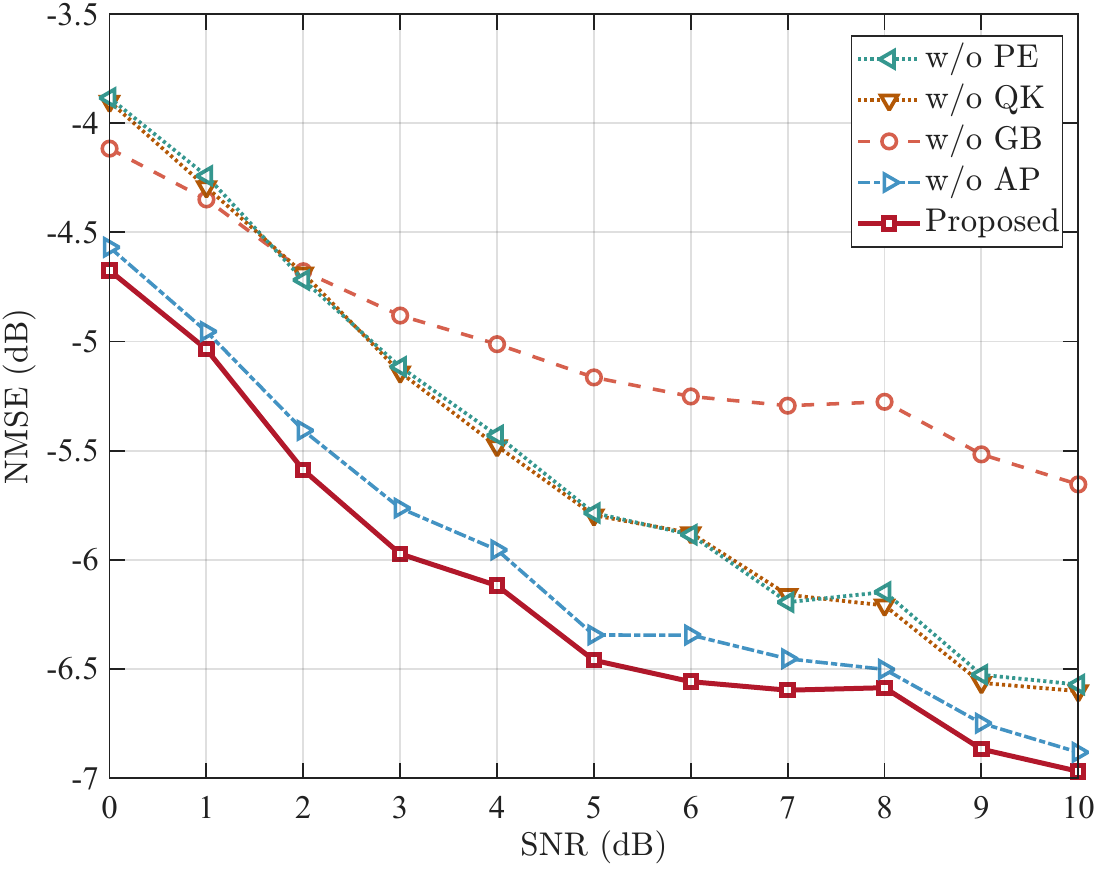} 
    \caption{CE NMSE versus SNR with fixed numbers of paths and pilot measurements, where the other parameters are set as $L=3$ and $Q=12$, respectively.}
    \label{fig:nmse_snr_ablation}
    
    \vspace{0.3cm}

    \includegraphics[height=0.65\columnwidth]{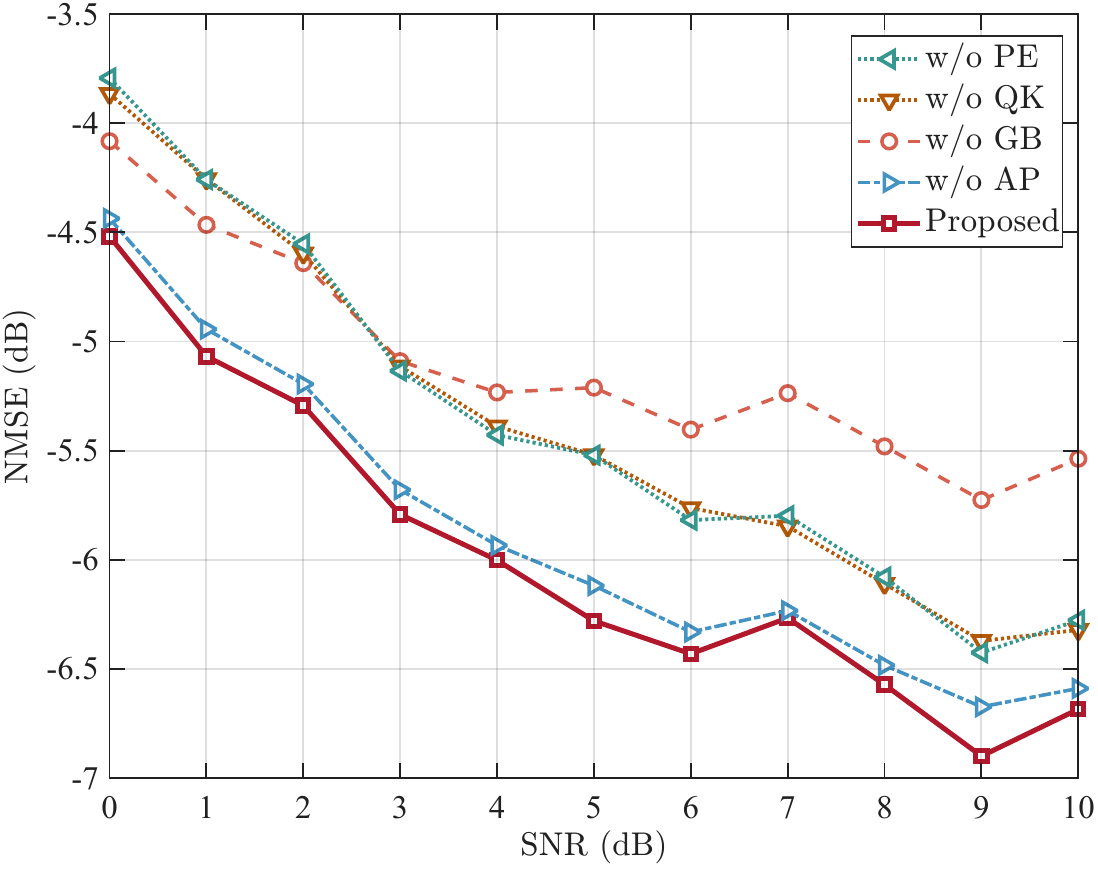} 
    \caption{CE NMSE versus SNR with random numbers of paths and pilot measurements, where $L$ and $Q$ are randomly generated for each channel realization.}
    \label{fig:nmse_snr_ablation_random}
\end{figure}

Finally, to validate the effectiveness of the key designs introduced in Section~\ref{sec: learning}, we conduct ablation studies by removing or replacing several components. Specifically, four variants are considered. 1) \textbf{Proposed (w/o GB)}: The Gram bias (GB) term in \eqref{eq:head_output} is removed, so that the output of the $m$-th head reduces to $\mathbf{H}_m^{(t)}=\mathrm{softmax}(\frac{\mathbf{Q}_m^{(t)}(\mathbf{K}_m^{(t)})^{\mathrm{T}}}{\sqrt{d_{\mathrm{head}}}})\mathbf{V}_m^{(t)}$. 2) \textbf{Proposed (w/o QK)}: The data-driven query-key attention term in \eqref{eq:head_output} is removed, and the attention map is determined only by the GB, yielding
$\mathbf{H}_m^{(t)}=\mathrm{softmax}([\mathbf{B}_{\mathbf{G}}]_{m,:,:})\mathbf{V}_m^{(t)}$. 3) \textbf{Proposed (w/o PE)}: The 2D PE along the angular and subarray dimensions, i.e., $\bar{\mathbf{E}}_{\text{ang}}$ and $\bar{\mathbf{E}}_{\text{sub}}$, are removed from $\mathbf{X}^{(0)}$. 4) \textbf{Proposed (w/o AP)}: The AP mechanism in \eqref{eq:angle_aggregation} is replaced by standard mean pooling along the subarray dimension. The support detection metrics are summarized in Table~\ref{tab:ablation}, while the corresponding CE NMSE performance under fixed and random channel configurations are illustrated in Fig.~\ref{fig:nmse_snr_ablation} and Fig.~\ref{fig:nmse_snr_ablation_random}, respectively.

As shown in Table~\ref{tab:ablation}, Fig.~\ref{fig:nmse_snr_ablation}, and Fig.~\ref{fig:nmse_snr_ablation_random}, removing the Gram-induced physical prior (w/o GB) leads to the most severe degradation, resulting in a $15.61\%$ drop in F1-score and a massive NMSE performance gap across all SNR regions. This highlights the vital role of the Gram prior in purifying high-dimensional representation of each grid. Deactivating query-key interactions (w/o QK) or PE (w/o PE) also compromises the F1-score (by $6.23\%$ and $5.38\%$) and noticeably elevates the NMSE, confirming that data-driven feature interaction and spatial geometry retention are both essential. Finally, removing the AP (w/o AP) incurs a negligible performance drop in the support detection. However, the full proposed model consistently achieves better channel reconstruction performance across the entire SNR range, demonstrating that adaptive AP refines group-level features and further improves the CE accuracy.

\begin{table}[t]
\caption{Ablation Study on the Test Set.}
\label{tab:ablation}
\centering
\renewcommand{\arraystretch}{1.2}
\begin{tabular}{@{}l c c c c@{}}
\toprule
\textbf{Variant} & \textbf{Prec. (\%)} & \textbf{Rec. (\%)} & \textbf{F1 (\%)} & \textbf{$\Delta$F1 (\%)} \\
\midrule
Proposed (w/o GB) & 81.03 & 62.75 & 70.73 & -15.61 \\
Proposed (w/o QK) & 86.07 & 74.91 & 80.11 & -6.23 \\
Proposed (w/o PE) & 87.51 & 75.33 & 80.96 & -5.38 \\
Proposed (w/o AP) & 91.87 & 81.29 & 86.26 & -0.08 \\
\midrule
\textbf{Proposed} & \textbf{91.29} & \textbf{81.89} & \textbf{86.34} & 0.00 \\
\bottomrule
\end{tabular}
\end{table}

\section{Conclusion}
\label{sec: conclusion}
This paper investigated spatially non-stationary CE for XL-MIMO systems from the perspective of sparse recovery. To overcome the model mismatch of the conventional DFT codebook, we proposed a novel JAS codebook that jointly characterizes the angular direction and the subarray-level visibility. Based on this representation, we further developed a signal processing and deep learning co-design framework, where CE was decomposed into Gram-attention-based parallel support detection followed by LS-based gain estimation. First, the matched filter response of each JAS grid was embedded into a high-dimensional feature vector. The Gram matrix was incorporated into the MHSA mechanism as a physical prior to guide data-driven feature refinement. The purified features were then utilized for hierarchical support prediction through decoupled heads, and the detected supports enabled LS-based gain estimation over a reconstructed low-dimensional dictionary. Simulation results showed that the proposed framework achieved superior support detection accuracy and CE NMSE compared with representative model-based and learning-based baselines. Ablation studies further verified the effectiveness of the Gram-domain physical prior and adaptive feature refinement. Overall, the proposed framework provides an effective learning-based solution for parallel support inference over a sparse representation. Future work may extend the proposed visibility-aware representation to uniform planar arrays (UPAs) and near-field channels by incorporating the elevation and distance dimensions, respectively, so that 2D angular structure, spherical wavefronts, and spatial non-stationarity can be jointly characterized.

\bibliographystyle{IEEEtran}
\bibliography{IEEEabrv,references}
\end{document}